\documentclass[preprint]{aastex}
\def\bm#1{\mbox{\boldmath $#1$}} 

\newcommand{\comments}[1]{}
\newcommand{\ba}{\begin{eqnarray}}
\newcommand{\ea}{\end{eqnarray}}
\newcommand{\be}{\begin{equation}}
\newcommand{\ee}{\end{equation}}
\newcommand{\lan}{\langle}
\newcommand{\ran}{\rangle}
\newcommand{\grad}{\nabla}

\begin{document}

\title{Thermal Instability with Anisotropic Thermal Conduction and Adiabatic Cosmic Rays: Implications for Cold Filaments in Galaxy Clusters}

\author{Prateek Sharma\altaffilmark{1}, Ian
  J. Parrish\altaffilmark{1}, Eliot Quataert} \affil{Astronomy
  Department and Theoretical Astrophysics Center, University of
  California, Berkeley, CA 94720} \email{psharma@astro.berkeley.edu, iparrish@astro.berkeley.edu, eliot@astro.berkeley.edu}

\altaffiltext{1}{Chandra/Einstein Fellow}
\begin{abstract}
  Observations of the cores of nearby galaxy clusters show H$\alpha$
  and molecular emission line filaments. We argue that these are the
  result of {\em local} thermal instability in a {\em globally} stable
  galaxy cluster core.  We present local, high resolution,
  two-dimensional magnetohydrodynamic simulations of thermal
  instability for conditions appropriate to the intracluster medium
  (ICM); the simulations include anisotropic thermal conduction along
  magnetic field lines and adiabatic cosmic rays.  Thermal conduction
  suppresses thermal instability along magnetic field lines on scales
  smaller than the Field length ($\gtrsim$10 kpc for the hot, diffuse
  ICM).  We show that the Field length in the cold medium must be
  resolved both along and perpendicular to the magnetic field in order
  to obtain numerically converged results.  Because of negligible
  conduction perpendicular to the magnetic field, thermal instability
  leads to fine scale structure in the perpendicular direction.
  Filaments of cold gas along magnetic field lines are thus a natural
  consequence of thermal instability with anisotropic thermal
  conduction. This is true even in the fully nonlinear regime and even
  for dynamically weak magnetic fields.  The filamentary structure in
  the cold gas is also imprinted on the diffuse X-ray emitting plasma
  in the neighboring hot ICM.  Nonlinearly, filaments of cold ($\sim
  10^4$ K) gas should have lengths (along the magnetic field)
  comparable to the Field length in the cold medium $\sim 10^{-4}$ pc!
  Observations show, however, that the atomic filaments in clusters
  are far more extended, $\sim 10$ kpc.  Cosmic ray pressure support
  (or a small scale turbulent magnetic pressure) may resolve this
  discrepancy: even a small cosmic ray pressure in the diffuse ICM,
  $\sim 10^{-4}$ of the thermal pressure, can be adiabatically
  compressed to provide significant pressure support in cold
  filaments.  This is qualitatively consistent with the large
  population of cosmic rays invoked to explain the atomic and
  molecular line ratios observed in filaments.
  
\end{abstract}

\section{Introduction}

The thermal instability has been studied extensively in the context of
the interstellar medium (ISM;
\citealt{fie65,koy00,san02,kri02,pio04,aud05}) and the formation of
solar prominences \citep[e.g.,][]{kar88}, but its role in galaxy
clusters has not received as much attention.  The cooling time in the
intracluster medium (ICM) near the centers of galaxy clusters may be
as short as 10-100 Myr. Observations show that there is a dramatic
{\em lack} of plasma below $\sim 1$ keV \citep[e.g.,][]{pet03},
inconsistent with the prediction of the original cooling flow models
\citep[e.g.,][]{fab94}. In addition, the star formation rate in the
central galaxy is 10--100 times smaller than if the gas cooled at the
predicted rate \citep[e.g.,][]{ode08}. This implies that cooling in
the intracluster medium (ICM) is balanced by some form of heating that
maintains an approximate global thermal equilibrium.  Feedback from a
central Active Galactic Nucleus (AGN) is an energetically plausible
source of the required heating (e.g., \citealt{guo08}). However,
precisely how the AGN provides this heating is not understood in
detail; nor are other heating mechanisms ruled out.  Many clusters
with a short central cooling time ($\lesssim$ 1 Gyr; or equivalently a
low central entropy) show both star formation and H$\alpha$ emission,
indicative of cool plasma at $\lesssim 10^4$ K \citep[e.g.,][]{cav08}.
Even if heating balances cooling in a global sense, the ICM plasma is
expected to be {\em locally} thermally unstable because of the form of
the cooling function \citep[e.g.,][]{fie65}. This local thermal
instability is an attractive mechanism for producing the H$\alpha$ and
molecular filaments seen in clusters with short cooling times.

Recently, the atomic and molecular filaments in the core of the
Perseus cluster have been spatially resolved
\citep[e.g.,][]{con01,sal06}. Based on the narrowness and coherence of
these filaments, \citet{fab08} suggested that magnetic fields play a
critical role in the dynamics of the filaments.  In addition to the
possible role of magnetic pressure and tension, the magnetic field also modifies
the microscopic transport processes in the ICM because the mean free
path along magnetic field lines is orders of magnitude larger than the
gyroradius.  As a result, thermal conduction is primarily along
magnetic field lines \citep[][]{bra65}.

This paper centers on carrying out magnetohydrodynamic (MHD)
simulations of thermal instability with thermal conduction along
magnetic field lines. We focus on understanding the physics of the
thermal instability in the ICM, rather than on making detailed
comparisons with observations.  Throughout this paper we ignore the
possible presence of a background gravitational field.  This allows us
to study the physics of the thermal instability without the added
complication of buoyant motions, inflow, etc.  In addition to
including the effects of anisotropic heat transport, we also include
cosmic rays as a second fluid.  Even an initially small cosmic ray
pressure can become energetically important in cold filaments. Part of
the motivation for including cosmic rays is that a significant
population of energetic ions ($\gg$ eV, the temperature of optical
filaments) appear required to explain the observed line ratios in
filaments in galaxy clusters \citep[][]{fer09}.

The multiphase nature of the ICM is physically analogous to the
well-studied multiphase ISM. The ISM has three dominant phases: a
molecular phase at $\sim$ 100 K, an atomic phase at $\sim 10^4$ K, and
the hot phase at $10^6$ K. The cooling function in the ISM is
thermally bistable with thermally stable phases at $\sim 100$ K and
$\sim 10^4$ K. The hot phase is thermally unstable but is probably
maintained at its temperature by supernova heating
\citep[][]{mck77}. The same physical considerations apply for the ICM,
except that the hot phase is maintained by a still poorly understood
heating process (e.g., AGN feedback).

This paper is organized as follows. Section 2 summarizes our model
equations and the results of a linear stability analysis including
conduction along field lines, cosmic rays, and magnetic fields (see
appendix A). Section 3 presents the numerical set-up and the results
of our numerical simulations.  Section 4 discusses the astrophysical
implications of our results.

\section{Governing Equations and Numerical Methods}
A magnetized plasma with cosmic rays can be described by the following
two-fluid equations:

\begin{equation}
\frac{d \rho}{d t} = - \rho \grad \cdot \bm{v},
\label{eq:cont}
\end{equation}
\begin{equation}
\rho \frac{d\bm{v}}{dt} = - \grad (p + p_{\rm cr}+\frac{B^2}{8\pi}) + \frac{(\bm{B}\cdot \grad)\bm{B}}{4\pi},
\label{eq:momentum}
\end{equation}
\begin{equation}
\frac{\partial \bm{B}}{\partial t} = \grad \times (\bm{v} \times \bm{B}),
\label{eq:ind}
\end{equation}
\begin{equation}
\frac{de}{dt} - \frac{\gamma e}{\rho} \,\frac{d\rho}{dt} 
= -n_en_i \Lambda(T) -\grad \cdot \bm{Q} + H(t),
\label{eq:energy}
\end{equation}
and
\begin{equation}
\frac{dp_{\rm cr}}{dt} - \frac{\gamma_{\rm cr} p_{\rm cr}}{\rho} 
\,\frac{d\rho}{dt} 
= - \grad \cdot \bm{\Gamma},
\label{eq:crenergy}
\end{equation}
where $d/dt = \partial/\partial t +
\bm{v} \cdot \grad$ is the Lagrangian time derivative, $\Lambda(T)$ is the cooling function,
\begin{equation}
\bm{Q} = - \kappa_\parallel \bm{ \hat{b}(\hat{b} \cdot \grad}) T
\label{eq:defQ}
\end{equation}
is the heat flux along magnetic field lines,
\begin{equation}
\bm{\Gamma} = - D_\parallel \bm{\hat{b}(\hat{b}} \cdot \grad) p_{\rm cr} 
\label{eq:defGamma}
\end{equation}
is the diffusive cosmic-ray energy flux (multiplied by $[\gamma_{\rm
  cr}-1]$), $\rho$ is the mass density, $n_e$ and $n_i$ are the
electron and ion number densities respectively, $\bm{v}$ is the common
bulk-flow velocity of the thermal plasma and cosmic rays, $\bm{B}$ is
the magnetic field, $\bm{\hat{b}} = \bm{B}/B$, $p$ and $p_{\rm cr}$
are the thermal-plasma and cosmic-ray pressures, $\kappa_\parallel$ is
the parallel thermal conductivity, $D_\parallel$ is the diffusion
coefficient for cosmic-ray transport along the magnetic field, and
$\gamma =5/3$ and $\gamma_{\rm cr}=4/3$ are the adiabatic indices of
the thermal plasma and cosmic rays, respectively.  We assume one-third
solar metallicity so that the mean molecular weights are $\mu=0.62$
and $\mu_e=1.18$.  We do not include gravity in the momentum equation
(Equation (\ref{eq:momentum})) in order to focus on the thermal
physics.  Note that our model equations also do not include the
streaming of cosmic rays relative to the thermal plasma, which
provides a mechanism for heating the thermal plasma (e.g.,
\citealt{loe91,guo08a}). This is numerically subtle to include
\citep{sha09a} and will be studied in future work.

It is difficult to study the problem of thermal instability without a
well-defined equilibrium state.  In order to ensure that we have such
a state, at each time step the heating term $H(t)$ in equation
(\ref{eq:energy}) is updated so that the volume averaged heating and
cooling in our computational domain balance each other. Without such
heating, the plasma as a whole cools to very low temperatures on a
cooling time (the same timescale on which the thermal instability is
developing).  Since the source of heating and its functional form are
not that well-understood in the ICM, we choose a constant heating per
unit volume for simplicity.  Calculations with a constant heating per
unit mass, i.e., $H(t) \propto \rho$, yield very similar results
because cooling ($\propto n^2$) dominates in the cold phase and
heating dominates in the hot phase in both cases.

\subsection{Linear Stability}

In appendix A we study the linear thermal stability of a uniform
plasma with magnetic fields, cosmic rays, and thermal conduction along
magnetic field lines.  To isolate the physics of interest in this
paper, we focus on the ``condensation mode,'' i.e., the entropy mode.
This calculation is a straightforward generalization of previous
results (e.g., \citealt{fie65}), 
but we include it for completeness. Here we quote the final results.

When the cosmic ray and magnetic pressure are negligible compared to
the plasma pressure, and the cooling time $t_{\rm cool}$ is long
compared to the sound-crossing time,
the growth rate for the thermal
instability is given by \be \gamma = -\chi_\parallel k_\parallel^2 -
t_{\rm cool}^{-1} \frac{d \ln (\Lambda/T^2)}{d\ln T},
\label{eq:isobaric}
\ee where $t_{\rm cool}^{-1} \equiv (\gamma-1)n_en_i\Lambda/p$.  Note
that the thermal conductivity, $\kappa_\parallel$, in Equation
(\ref{eq:defQ}) is related to the diffusivity used here,
$\chi_\parallel$, by $\kappa_\parallel = n_e k_B\chi_\parallel$.  The
first term on the right hand side of Equation (\ref{eq:isobaric})
describes the conductive stabilization of modes with short wavelengths
parallel to the local magnetic field (large $k_\parallel$).  This
implies that the fastest growing modes will be elongated along the
magnetic field lines and hence filamentary.  The critical parallel
length-scale at which $\gamma = 0$ (the Field length) is given by: \be
\lambda_F \equiv 2 \pi \left[\frac{\chi_\parallel t_{\rm cool}}{d \ln
    (T^2/\Lambda)/d\ln T}\right]^{1/2}. \label{eq:lambdaF}
 \end{equation}
 
 \begin{figure}
\centering
\epsscale{0.5}
\plotone{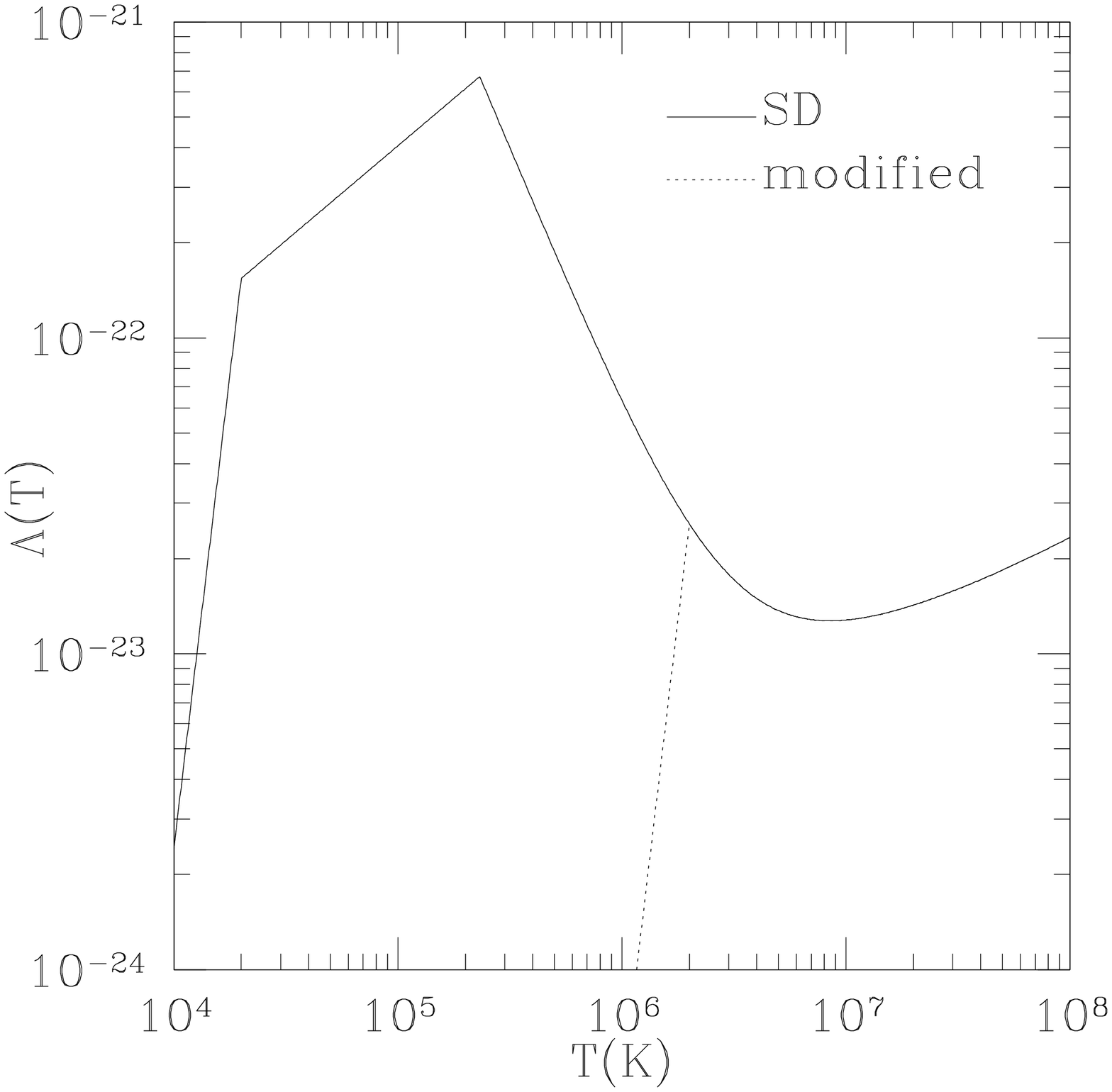}
\caption{Cooling function vs. temperature. Solid line: fit to the
  Sutherland \& Dopita cooling function for a third solar metallicity
  (Equation (\ref{eq:real_cf})).  Dotted line:   the modified cooling function 
  used in our simulations which has a stable cold phase at 
  $< 2\times10^6$ K.
  \label{fig:cf}}
\end{figure}

 When the cosmic ray and/or magnetic pressure is large compared to the
 plasma pressure, the isochoric growth rate applies, i.e., \be
\label{eq:isochoric}
\gamma = -\chi_\parallel k_\parallel^2 - t_{\rm cool}^{-1} \frac{d \ln
  \Lambda}{d\ln T}.  \ee This is also applicable when the cooling time
is shorter than the sound-crossing time, irrespective of the magnetic
and cosmic-ray contributions to the total pressure.  However, for
typical ICM conditions, the cooling time in the hot plasma is longer
than the sound-crossing time.

\subsection{Thermal Conductivity and the Cooling Function}
\label{sec:cond}

The thermal conductivity of a fully ionized plasma is governed by
electron collisions with the background ions and electrons. We are
interested in plasmas hotter than $10^{4}$ K, so that \be
\label{eq:spitzer}
\kappa_\parallel = \frac{1.84 \times 10^{-5}}{\ln \lambda} T^{5/2}
{\rm erg\; s^{-1} K^{-7/2} cm^{-1}}, \ee where we use a Coulomb
logarithm of $\ln \lambda = 37$. Since the Larmor radius of electrons
is much smaller than their collisional mean free path, thermal
conduction is only effective along magnetic field lines; the
perpendicular transport is negligible.  

We use the cooling function for an ionized plasma from
\citet{sut93}. A fit to the Sutherland and Dopita cooling rates for a
third solar metallicity is given by (a generalization of \citealt{toz01}): 
\ba \nonumber
\Lambda &=&  10^{-22} \, ( 8.6 \times 10^{-3} T_{\rm keV}^{-1.7} + 0.058 T_{\rm keV}^{0.5} + 0.063) \ {\rm ergs \, s^{-1} \, cm^{3}} \\
\nonumber
&&  {\rm~~~~~~~~~~~for~}T> 0.02~{\rm keV}, \\
\nonumber
\Lambda &=& 6.72\times 10^{-22} \, (T_{\rm keV}/0.02)^{0.6}  \ {\rm ergs \, s^{-1} \, cm^{3}} \\
\nonumber
&&  {\rm~~~~~~~~~~~for~} T \le 0.02 {\rm~keV, ~} T \ge 0.0017235 {\rm~keV}, \\
\nonumber
\Lambda &=& 1.544 \times 10^{-22} \, (T_{\rm keV}/0.0017235)^6  \ {\rm ergs \, s^{-1} \, cm^{3}} \\
\label{eq:real_cf}
&& {\rm~~~~~~~~~~~for~}T < 0.0017235~{\rm keV}, \ea where $T_{\rm
  keV}$ is the temperature in keV. For the cooling function given by Equation (\ref{eq:real_cf}), 
  the only phase that is thermally stable according to Equation
(\ref{eq:isobaric}) is plasma with $T \lesssim 0.0017$ keV $\simeq
10^4$ K.  

\citet{koy04} have shown that with isotropic conduction,
thermal instability simulations do not converge with increasing
resolution unless the Field length is always resolved by a few grid
cells.  We find the same result in our simulations.  The Field length
can be written as \be \lambda_F \approx 14.4 \ T_{\rm keV}^{7/4} \,
(n_{e,0.1}n_{i,0.1})^{-1/2} \, \Lambda_{-23}^{-1/2} \, \left[{d \ln
    (T^2/\Lambda) \over d\ln T}\right]^{-1/2} \, {\rm
  kpc} \label{eq:lf} \ee where $n_{e,0.1}$ ($n_{i,0.1}$) is the
electron (ion) number density in units of $0.1$ cm$^{-3}$ and $\Lambda
= 10^{-23} \Lambda_{-23}$ erg s$^{-1}$ cm$^3$.  Because the Field
length decreases rapidly with decreasing temperature, it is
prohibitive to resolve the Field length in a numerical simulation if
the plasma has a wide range of temperatures. Indeed, at fixed
pressure, Equation (\ref{eq:lf}) implies that the Field length at a
few $10^4$ K is $ \approx 10^8$ times smaller than at 1 keV!  In order
to ensure that our simulations always resolve the properties of the
cold phase of the ICM, we use an artificial cooling function in which
the thermally stable phase is at a much higher temperature of $2
\times 10^6$ K.  Because our simulations begin with plasma at typical
ICM temperatures $\sim 10^7$ K, the modest range of temperatures on
the computational domain makes it feasible to always resolve the Field
length.  Figure \ref{fig:cf} shows a comparison of the true cooling
function from Equation (\ref{eq:real_cf}) (solid line) and our
modified cooling function (dotted line).

With anisotropic thermal conduction, we find that numerical
convergence requires resolving {\em both the parallel and
  perpendicular} Field lengths in the cold stable phase.  More
precisely, if the perpendicular Field length is not resolved, the
width of cold structures perpendicular to the local magnetic field
decreases with increasing resolution.  In order to ensure that there
are no spurious results due to unresolved structures at the grid
scale, we thus include a constant {\em isotropic} diffusivity of $3
\times 10^{26}$ cm$^2$s$^{-1}$, in addition to the parallel thermal
conductivity given by Equation (\ref{eq:spitzer}).  The perpendicular
diffusivity required for numerical convergence depends on resolution;
we have verified numerically that $\approx 3\times 10^{26}$
cm$^2$s$^{-1}$ is the minimum isotropic diffusivity required to obtain
converged results for our two-dimensional simulations presented in \S
\ref{2D}.  The thermal diffusivity along magnetic field lines (Equation
(\ref{eq:spitzer})) is $\approx 30$ times larger than the isotropic
diffusivity for our typical initial conditions.  Thus thermal
conduction is still primarily along the magnetic field, although the
perpendicular conductivity is orders of magnitude larger than the
microscopic value.  In \S\ref{sec:conv}, we discuss how this
artificially large perpendicular conductivity might affect the
conclusions drawn from our simulations.

\subsection{Numerical Simulations}
\label{numerics}

In this paper, we carry out local one and two dimensional numerical
simulations of thermal instability for conditions appropriate to the
ICM.  We use unstratified local patches of the ICM to isolate the
physics of the thermal instability, as opposed to the buoyancy
instabilities present in a stratified, conductive plasma \citep{bal00,q08}.  We
focus on two-dimensional simulations -- as opposed to
three-dimensional simulations -- because of the challenging numerical
requirement of resolving the Field length in the cold medium (both
parallel and perpendicular to magnetic field lines).  Our box size is
40 kpc, somewhat larger than the typical Field length in the hot
medium; we use periodic boundary conditions.

We use the publicly available ZEUS-MP code \citep[][]{hay06} to solve
the MHD equations. Thermal conduction along magnetic field lines is
treated explicitly, using the method of \citet{sha07} to prevent unphysical 
negative temperatures. Since the stable timestep for conduction is much 
smaller than the MHD timestep, thermal conduction is subcycled. 
The cooling and heating terms in Equation (\ref{eq:energy}) are combined at each grid point,
and internal energy is updated by a first order explicit
(semi-implicit) method if heating (cooling) dominates; this ensures
that the internal energy is always positive, irrespective of the
timestep. Since the typical cooling time is much longer than the
sound-crossing time across a grid cell, this first order accurate
treatment of cooling is sufficient.

\begin{figure}
\centering
\epsscale{1.1}
\plottwo{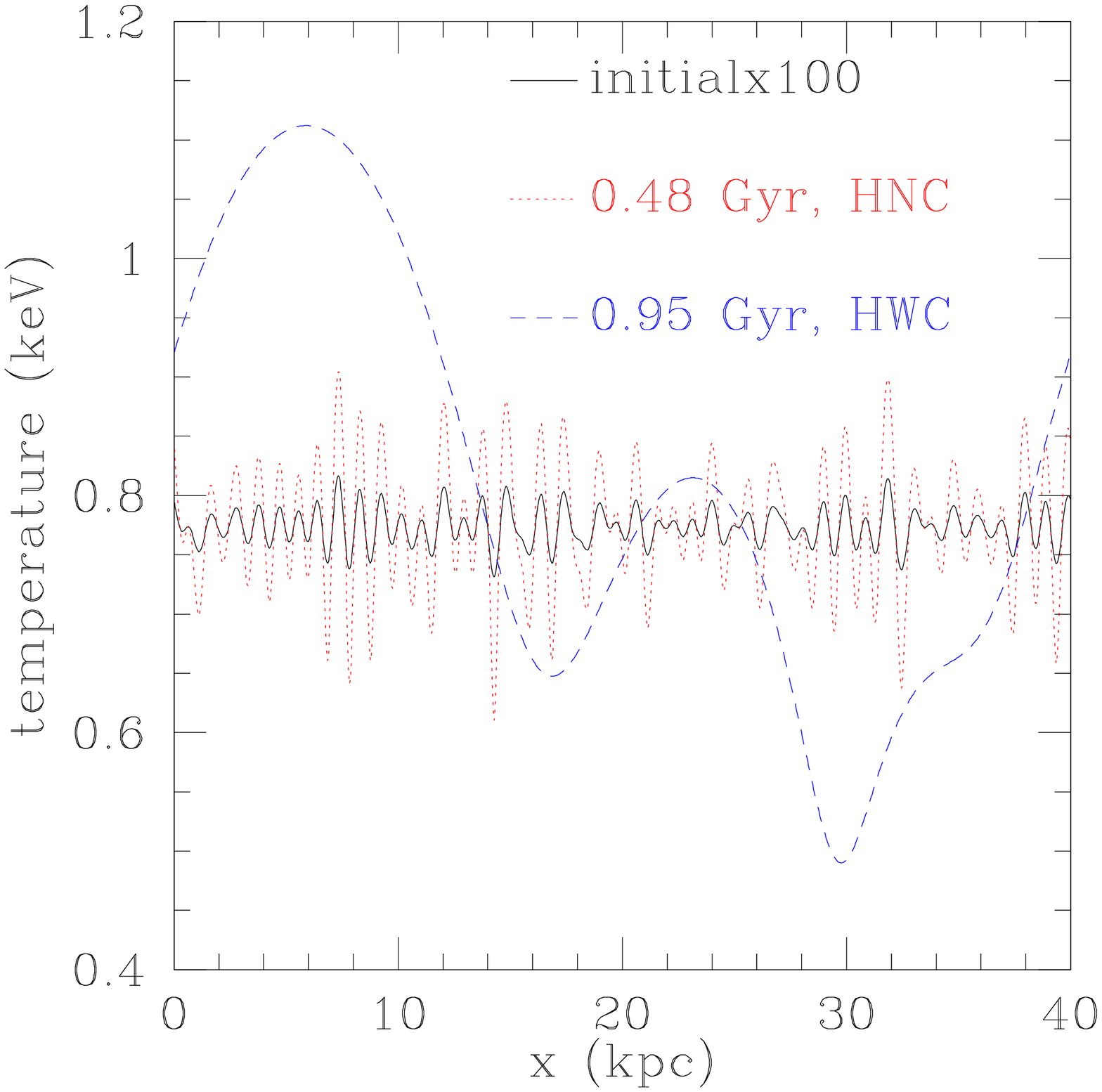}{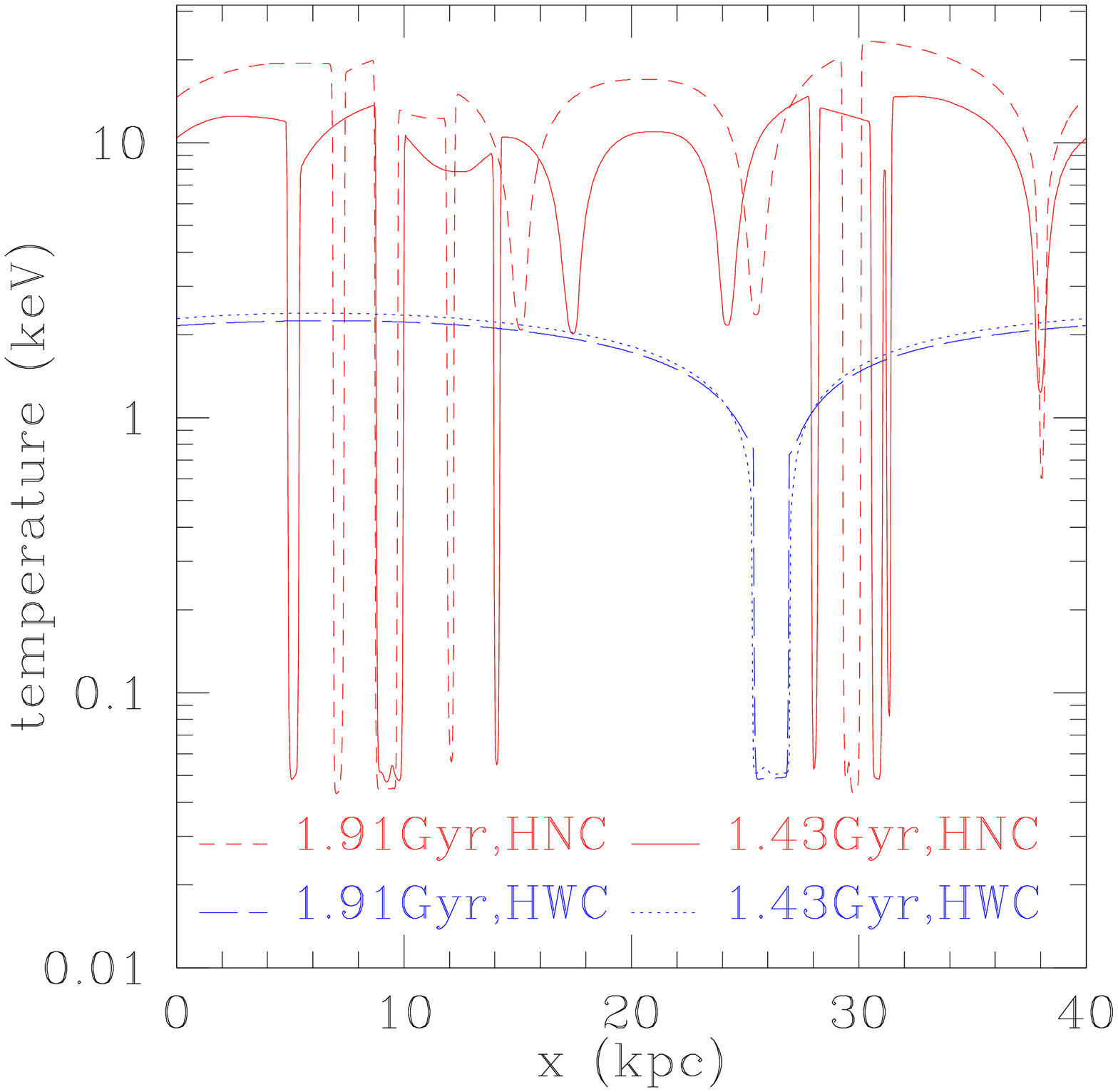}
\caption{Temperature profiles for one-dimensional runs with (HWC) and without (HNC) conduction  at different times in the linear
  ({\em left}) and nonlinear ({\em right}) regimes. The initial temperature
  fluctuations have been multiplied by a factor of 100 for clarity.
  \label{fig:1D}}
\end{figure}

Our initial condition consists of plasma with $T = 0.78$ keV and $n_e
= 0.1$ cm$^{-3}$; these parameters are characteristic of a reasonably
dense, low entropy ($\simeq 3.6$ keV cm$^2$) part of the ICM at small
radii, deep in the cluster core.  Note that the cooling time is longer
than the sound crossing time across the box so that the thermal
instability is in the roughly isobaric limit (except in
magnetic/cosmic ray dominated regions, where it behaves isochorically;
compare Equations (\ref{eq:isochoric}) \& (\ref{eq:isobaric})).  We initialize homogeneous and isotropic
$\sim 1 \%$ isobaric density/temperature perturbations on this initial equilibrium state;
the spectrum of initial perturbations is $\propto k$ for $k < k_0$ and
$\propto k^{-1}$ for $k > k_0$, so that most of the power is initially
at $\sim k_0$; $k_0$ corresponds to a scale $2\pi/k_0\approx$ 0.8 kpc 
for most of the simulations (we also experimented with smaller and larger $k_0$ for
comparison). The exact spectrum of initial perturbations is somewhat
arbitrary and is not well constrained in the ICM, but is also not
crucial for the subsequent evolution. All of the simulations with the
same $k_0$ have identical initial conditions and the initial
conditions are smoothed/interpolated for lower/higher resolution
simulations. This is important for quantitative testing of convergence
with increasing resolution.  The one-dimensional simulations do not
include magnetic fields, while the initial magnetic field is
$B=5~\mu$G and aligned at $45^0$ to the box in the two-dimensional
simulations. The simulations with cosmic rays begin with a uniform
cosmic ray pressure.

Tables \ref{tab:tab1} and \ref{tab:tab2} summarize the one and two
dimensional simulations discussed in detail in this paper. The tables
list the mass ($f_m$) and volume ($f_V$) fractions of plasma having a
temperature below $5\times 10^6$ K in the nonlinear state; this is a reasonable proxy for the
mass and volume fractions of the cold phase.  To quantify the spatial
coherence of the structures in the nonlinear state of the thermal
instability, we define the parallel and perpendicular length scales
of the density field via
\be L_\parallel \equiv \frac{ \int |\delta
  \rho| dV}{\int |\bm{\hat{b}\cdot\nabla} \delta \rho| dV}
\label{eq:Lpar}
\ee
and
\be
L_\perp\equiv \frac{\int |\delta \rho| dV}{\int |\bm{(\hat{z} \times \hat{b})\cdot \nabla} \delta \rho| dV},
\label{eq:Lperp}
\ee where $\delta \rho = \rho - \langle \rho \rangle$, $\langle \rho
\rangle$ is the volume averaged density (which is constant in time
because the mass in the computational domain is conserved), and
$\bm{\hat{z}}$ is perpendicular to the simulation plane. For the
one-dimensional simulations $\bm{\hat{x}}$ is used instead of
$\bm{\hat{b}}$ in Equation (\ref{eq:Lpar}) and $L_\perp$ is not
defined, so we only provide $L_\parallel$ in Table \ref{tab:tab1}.

\section{One-dimensional Simulations}
\label{sec:1d}

\begin{table}
\begin{center}
\caption{One dimensional runs \label{tab:tab1}}
\begin{tabular}{ccccccc}
  \tableline\tableline
  Label$^a$ & Res. & $\Delta x$ (kpc) & $L_\parallel$(kpc)$^b$ & $f_m^c$ & $f_V^c$ \\
  \tableline
  HWC & 1024  & 0.039 & 3.26 & 0.6  & 0.043 \\
  HNC & 1024 & 0.039 & 0.22 & 0.9 & 0.07 \\
  HWCl & 512  & 0.078 & 3.26 & 0.61 & 0.043  \\
  HWCh & 2048  & 0.02 & 3.26 & 0.6 & 0.042 \\
  HWCll & 256  & 0.16 & 0.89 & 0.67 & 0.043 \\
  \tableline
  \tablenotetext{a}{H stands for hydro. WC for with conduction. l \& h stand for lower and higher resolution runs. Initially $n_e=0.1$ cm$^{-3}$, $T=0.78$ keV, so that the cooling time is
    $\simeq 95$ Myr. The Field length in the initial condition is $\approx$ 10 kpc and at $2 \times 10^6$ K (temperature of the stable phase) is $\approx$ 0.07 kpc. 
    The box size is 40 kpc.}
  \tablenotetext{b}{$L_\parallel$ is defined in Equation (\ref{eq:Lpar}) and is evaluated at 1.43 Gyr, when the results have reached a quasi-steady state.}
  \tablenotetext{c}{$f_m$ ($f_V$) is the mass (volume) fraction of plasma below $5\times10^6$K (the ``cold phase'') evaluated at 1.43 Gyr.}
\end{tabular}
\end{center}
\end{table}
  
Figure \ref{fig:1D} shows temperature profiles in the linear ({\em
  left}) and nonlinear ({\em right}) regimes for one-dimensional
hydrodynamic simulations with (HWC) and without (HNC) thermal
conduction. For the simulation without conduction (dotted line) the
temperature fluctuations grow at all scales in the linear regime. By
contrast, for the run with conduction (dashed line) modes with scales
smaller than the Field length are suppressed by thermal conduction and
only the large scale modes grow.  Nonlinearly, the cold phase is
compressed into a smaller and smaller volume with time in the absence
of conduction, until the cold phase is unresolved (notice that
$L_\parallel$ in Table \ref{tab:tab1} is much smaller for the run
without conduction compared to the runs with conduction); the cold
peaks also merge, reducing the total number of dense peaks in time
(compare profiles at 1.43 and 1.91 Gyr for HNC in the {\em right} panel of Figure
\ref{fig:1D}).  Eventually all of the cold peaks will merge and
approach the grid scale because there is no heating of the cold phase
to prevent this. As the cold phase accumulates more and more mass in
time, the negligible mass in the hot phase becomes hotter and hotter
to conserve energy (which is enforced in our simulations via the
heating term $H(t)$ in Equation (\ref{eq:energy})).

\begin{figure}
\centering
\epsscale{0.5}
\plotone{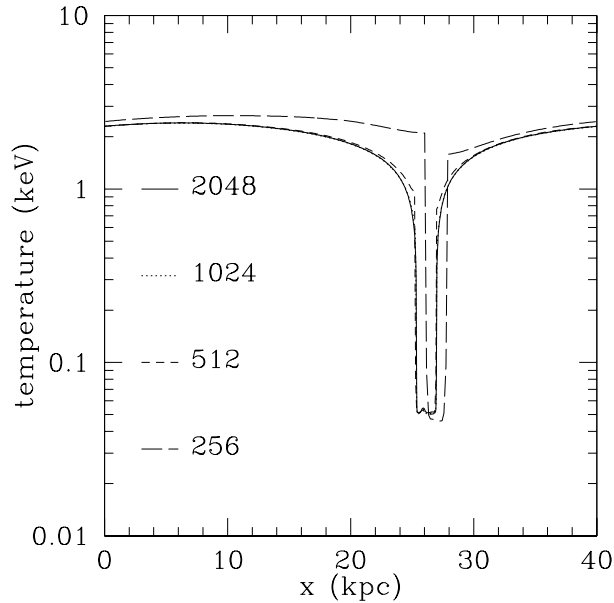}
\caption{Temperature profiles for one-dimensional simulations with conduction at $t=1.43$ Gyr 
for different resolutions: HWCll (256); HWCl (512); HWC (1024); and HWCh (2048).
 Convergence is achieved for $>$ 512  grid points.
 \label{fig:conv_1d}}
\end{figure}

The nonlinear evolution with conduction is qualitatively different:
there is only one cold region (this is because of the large Field
length in the initial plasma; this result is insensitive to the
initial density fluctuation spectrum) and the hot phase saturates at a
temperature $\approx 2$ keV, much cooler than in the simulations
without conduction.  The temperature of the cold phase is, however,
the same with and without conduction; this is set by the temperature
of the thermally stable branch of the cooling function. Figure
\ref{fig:1D} shows that in the presence of conduction, the temperature
profile reaches an approximate steady state, with very little change
from 1.43 to 1.91 Gyr.  The steady state requires both the additional
heating $H(t)$ in Equation (\ref{eq:energy}) and thermal conduction.
In particular, the cooling is dominated by the dense, cold gas while
the extra heating is primarily supplied to the hot phase (because $H$
is constant per unit volume).  This extra heating is conducted to the
rapidly cooling ($\propto n^2$) cold phase producing a steady
state. This energy transfer from the hot to the cold phase can only be
properly captured if the Field length in the cold phase is resolved,
which is why it is critical to do so to obtain converged results (see
\citealt{koy04}).

Figure \ref{fig:conv_1d} shows temperature profiles at 1.43 Gyr for
simulations including thermal conduction at several different
resolutions.  The temperature profile is reasonably converged only for
simulations with more than 512 grid points; in particular, note that
the physical size of the cold phase does not change with resolution
for $N>512$.  This is only true when the Field length in the cold
phase is resolved.  The Field length for the initial temperature and
density is $\approx 10$ kpc.  The Field length for the isobaric cold phase at $2 \times
10^6$ K (the stable phase for our modified cooling curve; Figure
\ref{fig:cf}) is $\approx 0.07$ kpc; this just
starts to be resolved at more than 512 grid points since our box size
is $40$ kpc, and thus $\Delta x=0.078$ kpc at $N = 512$.

\section{Two-dimensional Simulations}
\label{2D}

Having used one-dimensional simulations to describe the basic physics
of the thermal instability and the numerical requirements for
simulating it, we now turn to the more physically realistic case of
two-dimensional simulations. As we have emphasized previously, the
Field length must be resolved both parallel and perpendicular to the
local magnetic field in multi-dimensional simulations of the thermal
instability.  This is why we (1) include an isotropic thermal
conductivity (which helps resolve structures perpendicular to the
field; see \S \ref{sec:cond}) in addition to the parallel conductivity
and (2) artificially increase the temperature of the thermally stable
phase (Figure \ref{fig:cf}).

\begin{table}
\begin{center}
\caption{Two dimensional runs \label{tab:tab2}}
\begin{tabular}{ccccccccc}
\tableline\tableline
Label$^{a}$ & Res. & $\Delta x = \Delta y$ (kpc) & $p_{\rm cr}/p$ & $D_\parallel$ (cm$^2$s$^{-1}$) & $\frac{\lan {\rm heating} \ran }{\lan {\rm cooling} \ran}$ &  $L_\parallel/L_\perp^{b}$ & $f_m^c$ & $f_V^c$ \\
\tableline
MWC$^\star$ & 1024 & 0.039 & 0 & - & 1 & 2.48 & 0.51 & 0.061 \\
MWCl & 512 & 0.078 & 0 & - & 1 & 2.18 & 0.5 & 0.06 \\
MWCh & 2048  & 0.02 & 0 & - & 1 & 2.55 & 0.51 & 0.064 \\
MWIC & 1024 & 0.039 & 0 & - & 1 & 1.08 & 0.48 & 0.041 \\
MWCCR & 1024 & 0.039 & 0.1 & 0 & 1 & 3.45 & 0.41 & 0.12 \\
MWCCRs & 1024 & 0.039 & 10$^{-3}$ & 0 & 1 & 2.5 & 0.5 & 0.063 \\
MWCCRd28 & 1024 & 0.039 & 0.1 & $10^{28}$ & 1 & 2.66 & 0.4 & 0.1 \\
MWCCRd30 & 1024 & 0.039 & 0.1 & $10^{30}$ & 1 & 2.29  & 0.47 & 0.061 \\
MWCh0.9c & 1024 & 0.039 & 0 & - & 0.9 & 3.91 & 0.9 & 0.38 \\
MWCh1.05c & 1024 & 0.039 & 0 & - & 1.05 & 3.86 & 0.018 & $1.5\times 10^{-3}$\\
\tableline
\end{tabular}
\tablenotetext{a}{M stands for MHD. WC means with conduction, IC is
  for isotropic conduction, CR for cosmic rays.  l \& h stand for
  lower and higher resolution runs. All runs have a small isotropic
  conduction added for convergence (see \S \ref{sec:cond}).  Initially
  $n_e=0.1$ cm$^{-3}$ and $T=0.78$ keV, so that the cooling time is
  $\simeq 95$ Myr. Initial magnetic field is 5 $\mu$G and aligned $45^0$ 
  to the two-dimensional cartesian box.  The Field length in the initial condition is
  $\approx$ 10 kpc and at $2 \times 10^6$ K (temperature of the stable
  phase) is $\approx$ 0.07 kpc.  The box size is 40 kpc. Some less
  crucial simulations are not included in the table but are discussed
  in the text.}  \tablenotetext{\star}{The fiducial run.}
\tablenotetext{b}{$L_\parallel$ and $L_\perp$ are defined in
  Equations (\ref{eq:Lpar}) \& (\ref{eq:Lperp}), and are evaluated at 0.95
  Gyr.}  \tablenotetext{c}{$f_m$ ($f_V$) is the mass (volume)
  fraction of plasma below $5\times10^6$K (the ``cold phase'')
  evaluated at 0.95 Gyr (except for MWIC where these are evaluated at
  1.43 Gyr).}
\end{center}
\end{table}

\subsection{The Fiducial Run: MHD with Anisotropic Thermal Conduction}
\begin{figure*}
\centering
\epsscale{1.1}
\plotone{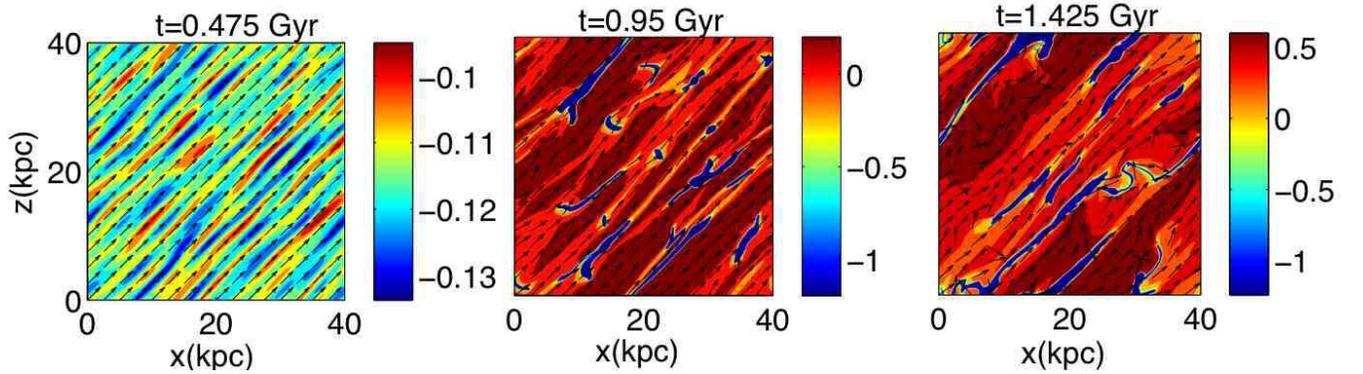}
\caption{Contour plots of $Log_{10}$ temperature (in keV) for the fiducial run (MWC) at linear (0.475 Gyr; {\em left}) and nonlinear 
(0.95 Gyr, {\em center}; 1.425 Gyr, {\em right}) stages of the instability. 
The arrows show the magnetic field direction.
\label{fig:2D}}
\end{figure*}

Our fiducial two-dimensional simulation is MWC summarized in Table
\ref{tab:tab2}; this is an MHD simulation with anisotropic thermal
conduction and an initial magnetic field of $B = 5 \mu$G aligned at
45$^0$ relative to the x-axis, in the plane of the simulation.  The
initial density (0.1 cm$^{-3}$) and temperature ($0.78$ keV)
correspond to an initial cooling time of 95 Myr and a Field length
$\approx 10$ kpc along the magnetic field and $\approx 1.7$ kpc
perpendicular to the field; the magnetic field initially contributes
only $\simeq 0.4 \%$ ($\beta \equiv 8\pi p/B^2 \simeq 250$) of 
the total pressure.  Figure \ref{fig:2D} shows
contour plots of the temperature in the linear (0.475 Gyr) and
nonlinear (0.95, 1.425 Gyr) regimes, along with arrows showing the magnetic
field direction at each time; because the pressure remains relatively
constant even nonlinearly, density scales nearly as the inverse of temperature. 

Figure \ref{fig:2D} shows that the thermal instability develops
anisotropically, with a filamentary structure along the magnetic
field; this is because thermal conduction efficiently suppresses
small-scale structures along the field, but not across it.
Quantitatively, the ratio $L_\parallel/L_\perp$ measures the
anisotropy of filaments with respect to the magnetic field; this is
$\sim 2.5$ at 0.95 Gyr (Table \ref{tab:tab2}) and increases  to $\sim 3.5$ at 
later times (see the {\em left} panel in Figure
\ref{fig:conv}).  The number of cold filaments decreases in time
because some of the filaments merge
together nonlinearly. 
Interestingly, the majority of the cold filaments are
oriented along the direction of the local magnetic field even in the
nonlinear regime. 
Some of the filaments at 0.95 Gyr are quite small and
relatively isotropic because of small and nearly isotropic conduction in the cold 
phase.
However, at later times (e.g., 1.425 Gyr) the small filaments coalesce 
to form large ones. The nonlinear development of the thermal instability proceeds 
in two phases: in the first phase nonlinear filaments aligned along field lines 
condense from the hot ICM, becoming shorter in time because of a smaller conductivity   
in the cold phase; in the second stage these cold filaments with large velocities (primarily
along themselves) merge to form longer filaments. This is clearly seen in Figure \ref{fig:conv}
as an increase in $L_\parallel/L_\perp$ after an initial dip at $\sim 1$ Gyr.

\begin{figure*}
\centering
\epsscale{1.}
\plotone{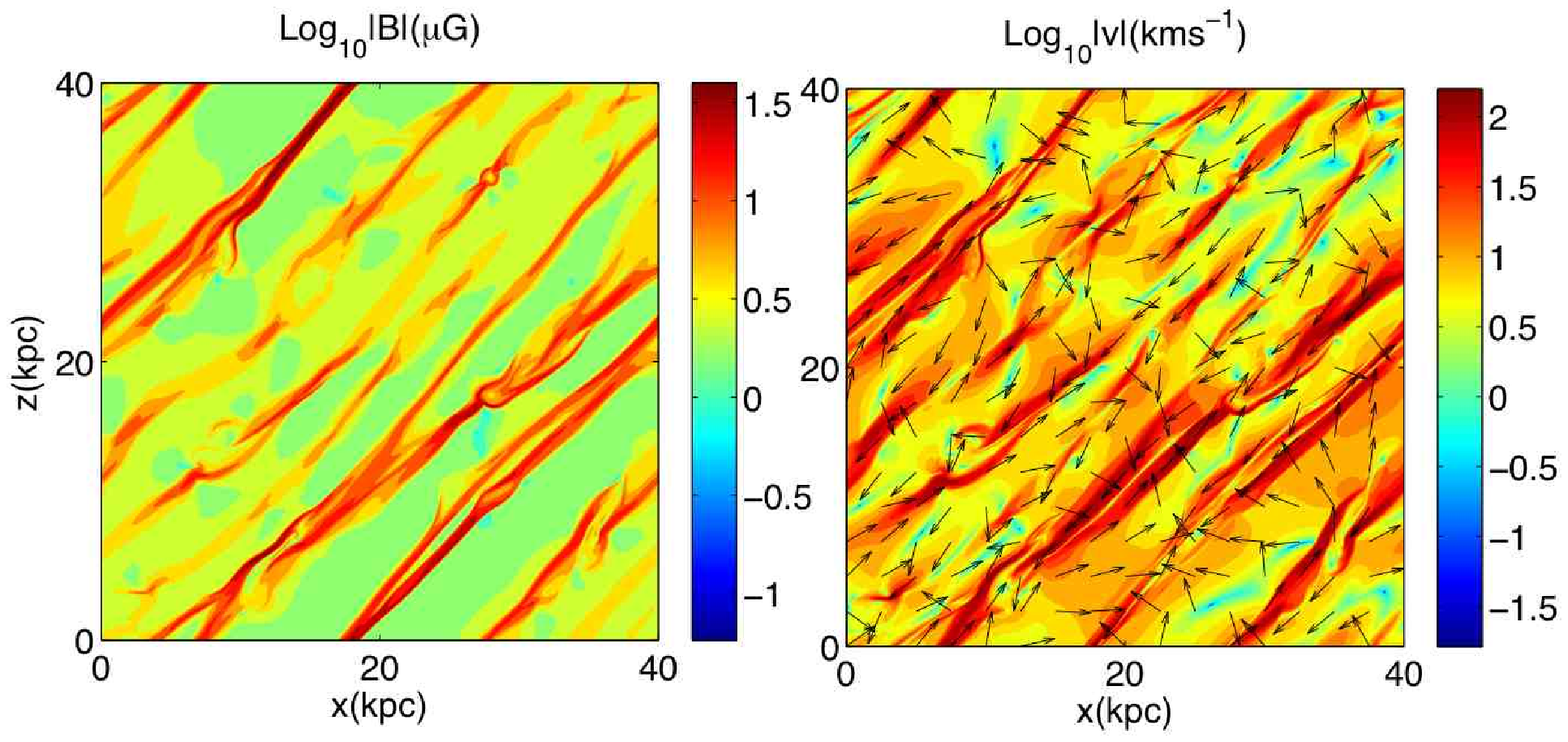}
\caption{Contour plots showing $Log_{10}|B|$ (magnitude of the magnetic
  field strength) ({\em left}) and $Log_{10}|v|$ (magnitude of the velocity) ({\em right}) for  the fiducial run at $0.95$ Gyr. The arrows in the velocity plot show  the direction of the velocity unit vector.
  \label{fig:2D_detail}}
\end{figure*}

Figure \ref{fig:2D} shows that the direction of the magnetic field is
only moderately perturbed from its initial direction even in the fully
nonlinear regime.  However, the magnetic field strength increases by a
factor of $\gtrsim 3-8$ in the cold filaments (see the {\em left}
panel of Figure \ref{fig:2D_detail}), to the point where the magnetic
pressure is important in the filaments.  The regions over which the
field is enhanced are coincident with, but significantly longer than,
the location of the cold filaments. The field enhancement occurs via
flux freezing as the cooling plasma is compressed perpendicular to the
initial field direction in the nonlinear state of the thermal
instability; analogous compression along the field lines is suppressed
because of thermal conduction.  In the hot diffuse gas between the
filaments, the magnetic field decreases by a factor $\simeq 2-3$ from
its initial value of $\approx$ 5 $\mu$G.  Note that for a realistic
cooling function, the density contrast between the filaments and the
diffuse medium will be larger than is found in our simulations, and so
the magnetic field compression in the filaments will also be stronger.

The {\em right} panel of Figure \ref{fig:2D_detail} shows that the
velocities driven by the thermal instability can reach $30-100$ km
s$^{-1}$, comparable to the sound speed in the cold filaments, but
much less than the sound speed in the hot phase.  Such high velocities
can disrupt the tendency of buoyancy instabilities 
in the hot phase of the ICM to reorient the magnetic field  \citep[e.g.,][]{sha09b,par09b}. 
The high velocities
are spatially coincident with the magnetic field enhancements and the
cold filaments. The velocity vectors generally point toward the cold
filaments in the hot phase, showing that mass from the hot thermally
unstable medium is condensing into the cold phase. This flow of mass
is, however, transient.  The thermal instability reaches a steady
state in which cooling from the dense, cool ICM is balanced by
conductive heating from the hot ICM, which is in turn heated
(artificially) by our external heat source $H(t)$ in Equation
(\ref{eq:energy}).  Once this steady state is established, mass flow
between the phases is significantly reduced. Although mass flow across
the phases is reduced, the cold filaments retain large velocities
along themselves and the volume averaged velocity is $\sim$ 20
kms$^{-1}$ (see the {\em right} panel of Figure  \ref{fig:conv}
discussed later).

\begin{figure}
\centering
\epsscale{1.}
\plottwo{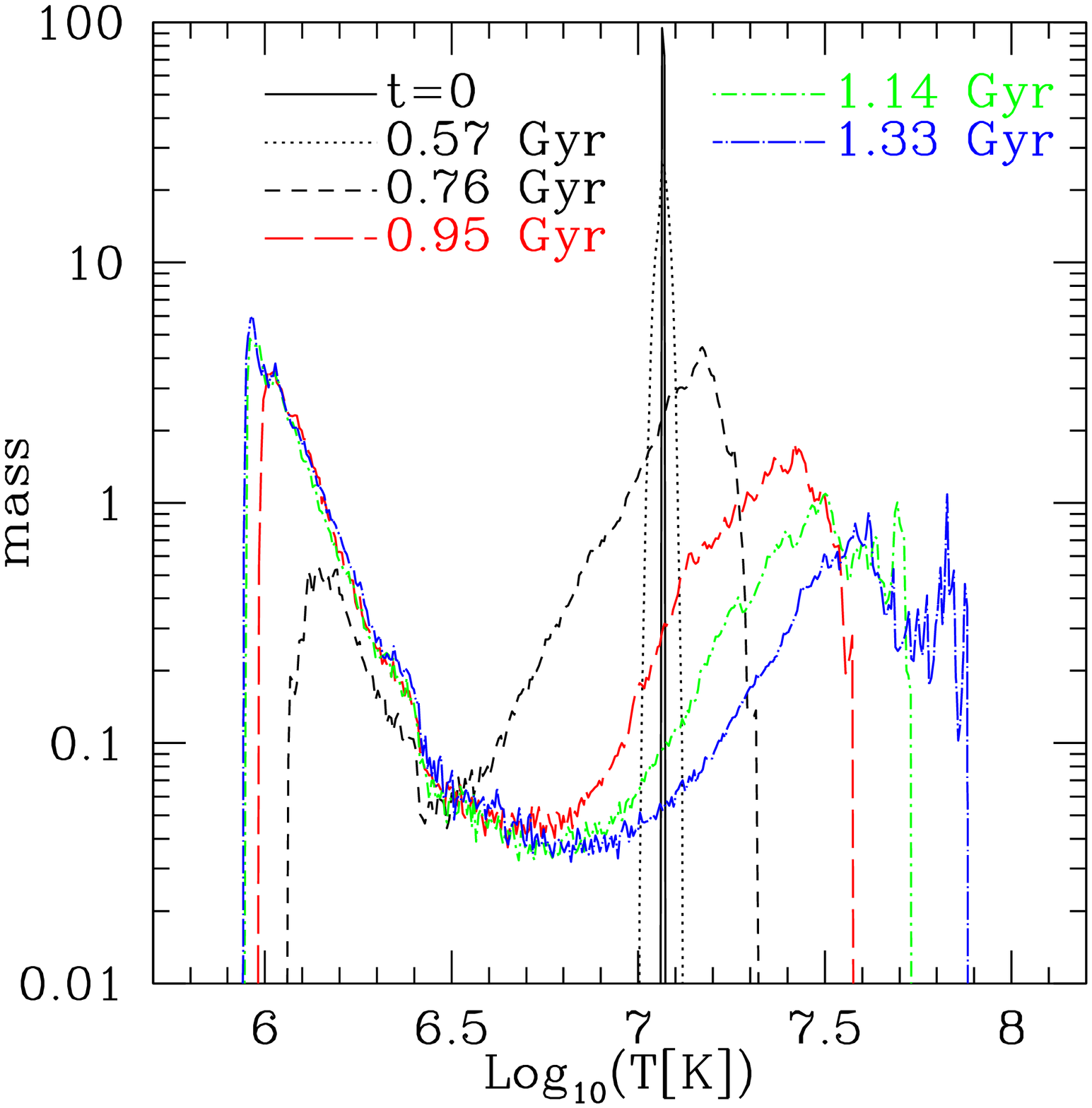}{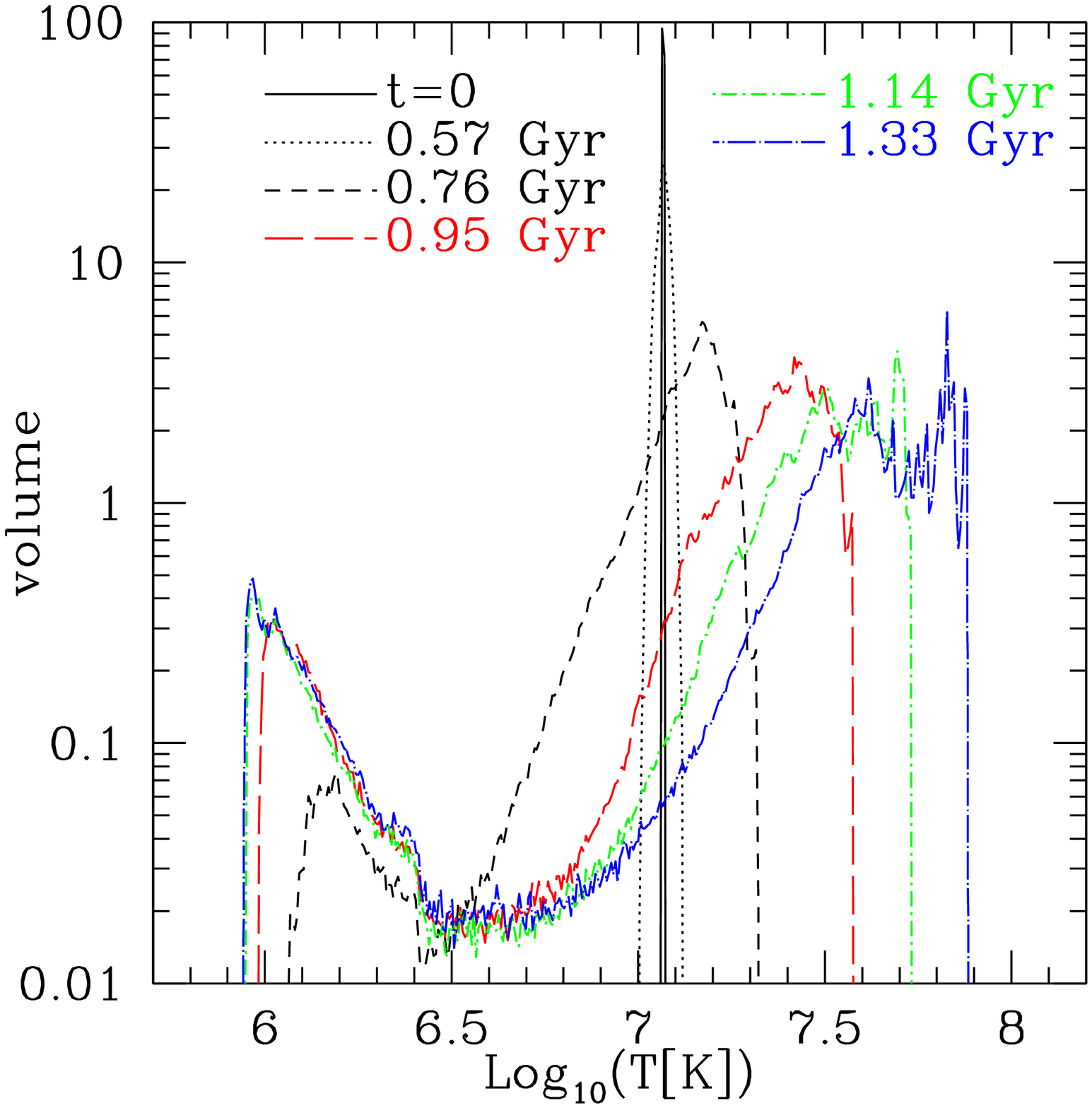}
\caption{The mass ($dM/d\log_{10}T$; {\em left}) and volume ( $dV/d\log_{10}T$; {\em right}) fractions occupied by plasma of a given temperature $T$ for the fiducial run (MWC) at different times.  The
normalization is such that the total mass/volume under the curve is unity. The initial cooling time is $\simeq 0.1$ Gyr and the simulations begin to saturate after $\simeq 0.8$ Gyr. The hottest plasma in the box  becomes hotter 
with time.
  \label{fig:phase_time}}
\end{figure}

Nonlinearly, the plasma exists in two phases, with very little plasma
at the intermediate temperatures.  Figure \ref{fig:phase_time} shows
the mass ({\em left} panel) and volume ({\em right} panel)
distribution of plasma at different times for the fiducial run. The
plasma is at $\approx 10^7$ K initially but evolves into a two-phase
structure. The phase structure evolves rapidly at early times (before
$\sim 1$ Gyr), but the evolution is slower at later times. The mass
and volume occupied by the plasma at intermediate temperatures
decreases in time. The ``mass dropout rate,'' (i.e., the rate at which
plasma cools below a given temperature) at $10^7$ K is large
initially, but once a two-phase medium is established, the mass and
volume of the hot and cold phases are roughly constant in time, with
very little mass dropout. While there is significant mass in the cold
filaments, most of the volume is occupied by the hot phase (see $f_m$
and $f_V$ in Table \ref{tab:tab2}). The hottest plasma in the domain
slowly becomes hotter with time in the two dimensional simulations; by
contrast, in 1D the plasma reaches a steady state at 1.43 Gyr (Figure
\ref{fig:1D}). It takes longer to reach a quasi-steady state in two
dimensions because it is easier for hot isothermal regions to become
thermally isolated from the cold plasma (because of the small
perpendicular conductivity). Since the hottest plasma becomes hotter 
with time and the conductivity is  a strong function of temperature 
(Equation (\ref{eq:spitzer})), it becomes difficult to run the simulations 
for long times.

\subsection{Simulations with Isotropic Thermal Conduction}

\begin{figure}
\centering
\epsscale{1.0}
\plotone{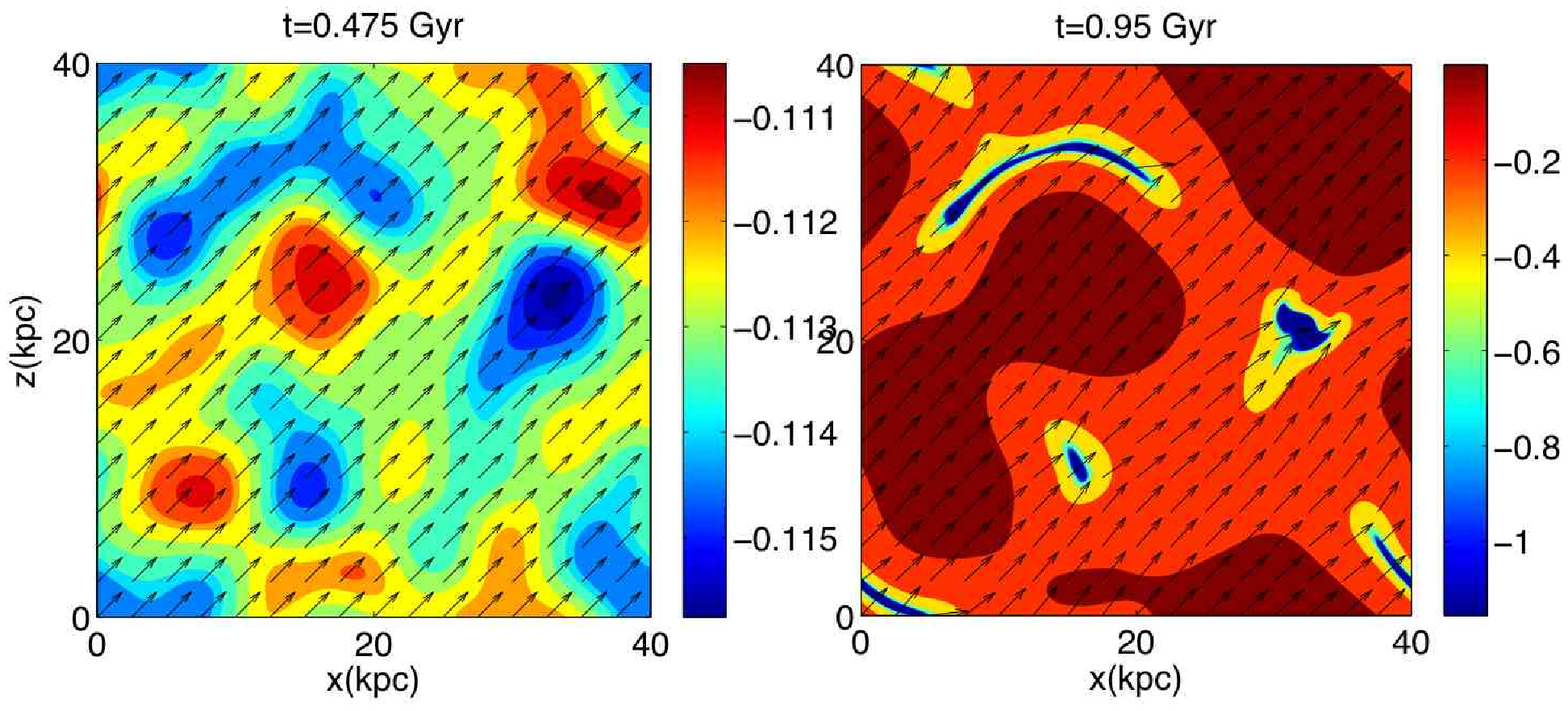}
\caption{Contour plots of $Log_{10}$ temperature (keV) for the simulation with isotropic thermal conduction at the Spitzer value (MWIC), at 0.475 Gyr ({\em left}) and 0.95 Gyr ({\em right}). The arrows show the magnetic field direction.
  \label{fig:isocond}}
\end{figure}

\begin{figure}
\centering
\epsscale{0.5}
\plotone{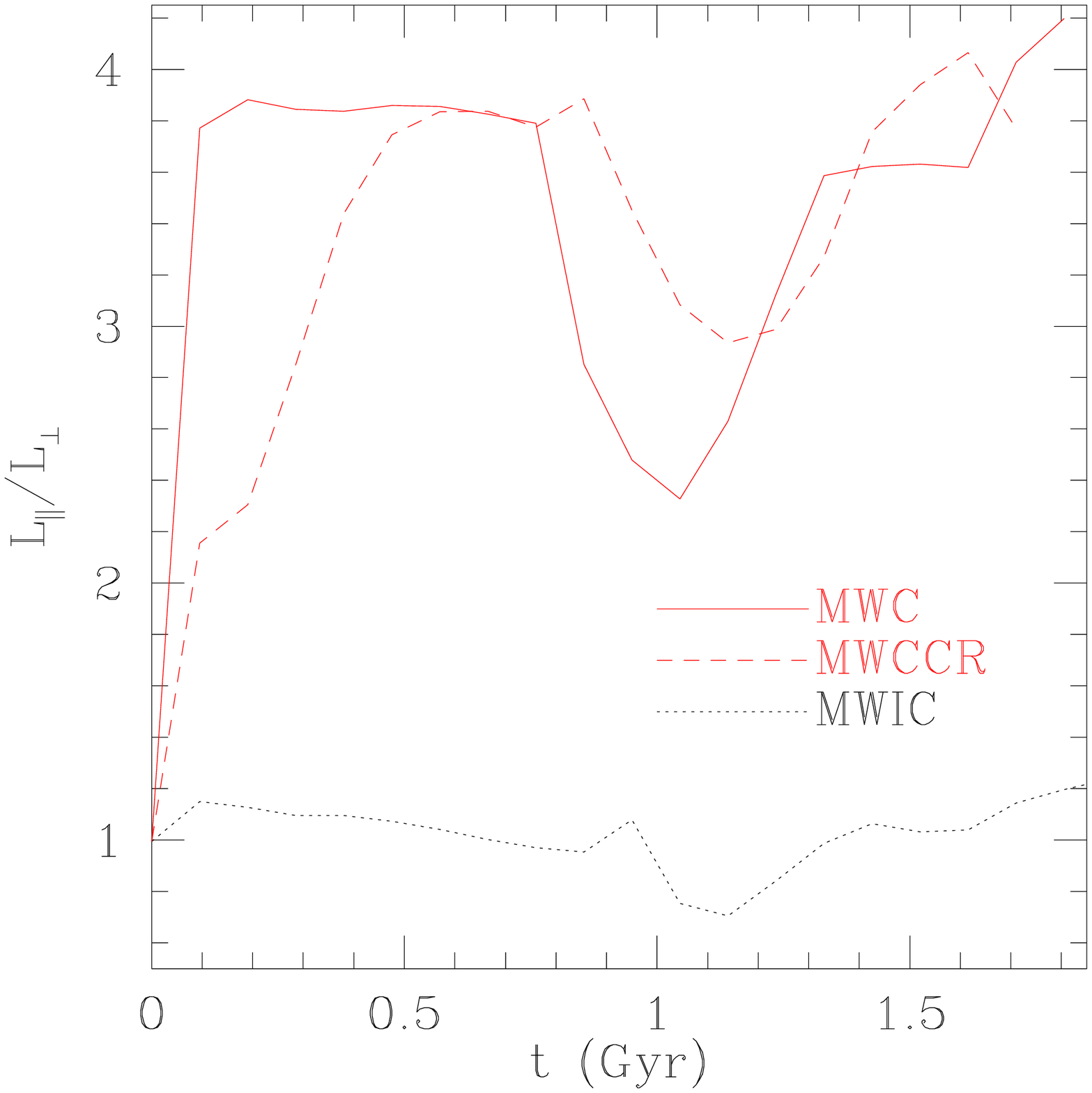}
\caption{The anisotropy of the density field $L_\parallel/L_\perp$ as a function of time for different runs: the fiducial run (MWC),  the run with initial $p_{\rm cr}/p=0.1$ (MWCCR), and the run with isotropic conduction (MWIC). Note also that the filaments are longer and broader for simulations that include 
cosmic rays, i.e., both $L_\parallel$ \& $L_\perp$ are larger even though $L_\parallel/L_\perp$ is comparable (see Figures \ref{fig:2D} \& \ref{fig:cosmic}).
\label{fig:lbyl_comp}}
\end{figure}

To assess the importance of including anisotropic thermal conduction,
we carried out simulations identical to the fiducial run in every way
except that the conductivity is isotropic at the Spitzer value (MWIC
in Table \ref{tab:tab2}).  Figure \ref{fig:isocond} shows the
temperature contour plots at 0.475 Gyr ({\em left} panel) and 0.95 Gyr
({\em right} panel). In the linear state the modes are isotropic and
on relatively large scales, irrespective of the magnetic field
direction. By contrast, with anisotropic conduction, the cold plasma
is filamentary even in the linear state (Figure
\ref{fig:2D}).\footnote{The Field length perpendicular to the magnetic
  field is much smaller in the simulation with anisotropic conduction
  than in the simulation with isotropic conduction.  This is why there
  is much more small-scale structure, and more cold `filaments,' in
  Figure \ref{fig:2D} than in Figure \ref{fig:isocond}.  In addition,
  because we initialize power primarily at $\approx 0.8$ kpc (\S
  \ref{numerics}), the amplitude of the initial perturbations that can
  actually grow ($\gtrsim$ the Field length) is larger in the
  simulation with anisotropic conduction.  These perturbations thus
  evolve somewhat more rapidly.}  Nonlinearly, the orientation of the
cold plasma in simulations with isotropic conduction is unrelated to
-- or even somewhat perpendicular to (see dotted line in Figure \ref{fig:lbyl_comp}) -- the local magnetic field
direction, unlike in simulations with anisotropic conduction, where
the filaments develop along the magnetic field (Figure \ref{fig:2D}).
Although the morphology of the cold gas is different in the two cases,
the evolution of the phase structure is qualitatively similar; there
is significant mass in the cold phase, but the volume is dominated by
the hot phase. The differences between Figures \ref{fig:2D} and
\ref{fig:isocond} emphasize the critical importance of including
anisotropic thermal conduction when studying the thermal physics of
galaxy cluster plasmas.

\subsection{Convergence of Two-dimensional Simulations}
\label{sec:conv}
\begin{figure}
\centering
\epsscale{1.}
\plotone{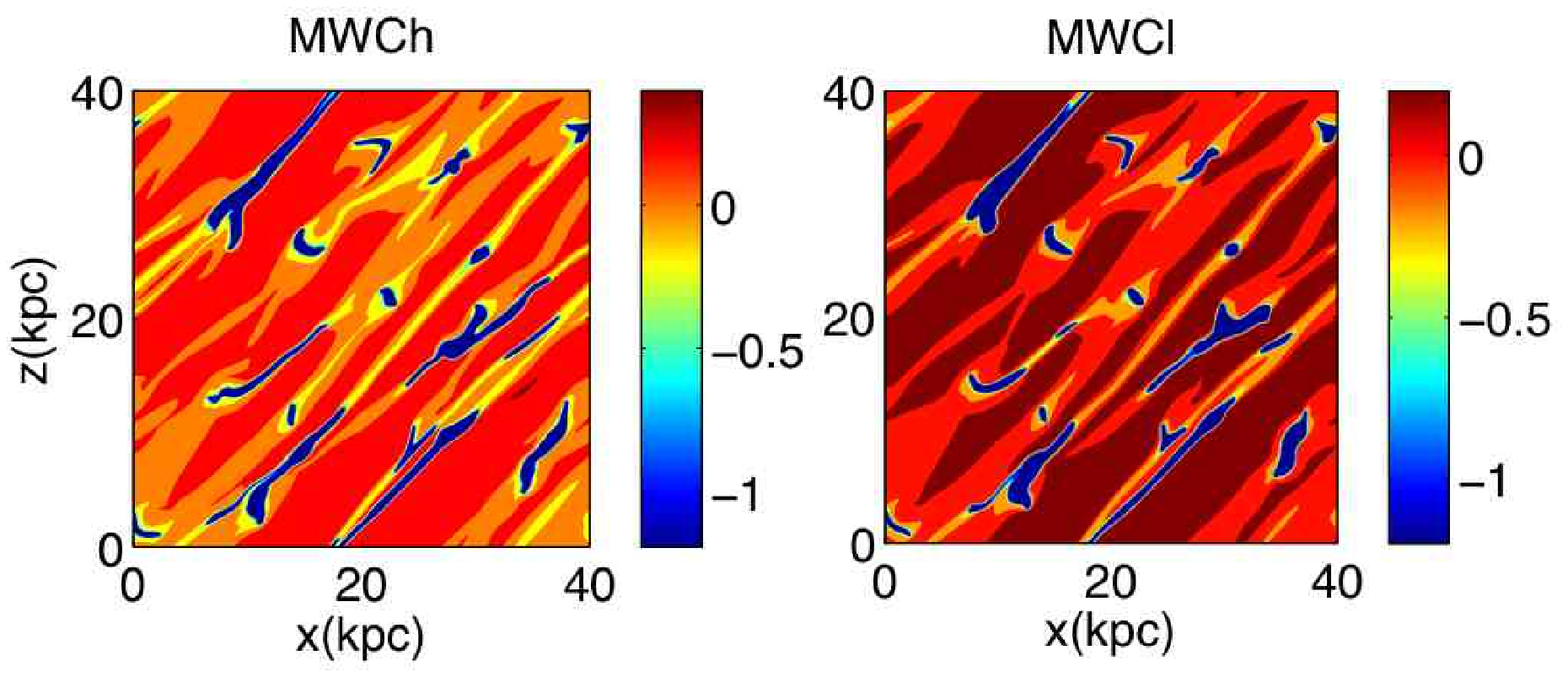}
\caption{Contour plots of $Log_{10}$ temperature (keV) at 0.95 Gyr for higher (MWCh; {\em left}) and lower (MWCl; {\em right}) resolution analogues of our fiducial simulation.  Figure \ref{fig:2D} shows the corresponding temperature plot for the fiducial run.   All three are reasonably similar.
  \label{fig:conv_2d}}
\end{figure}
\begin{figure}
\centering
\epsscale{1.}
\plotone{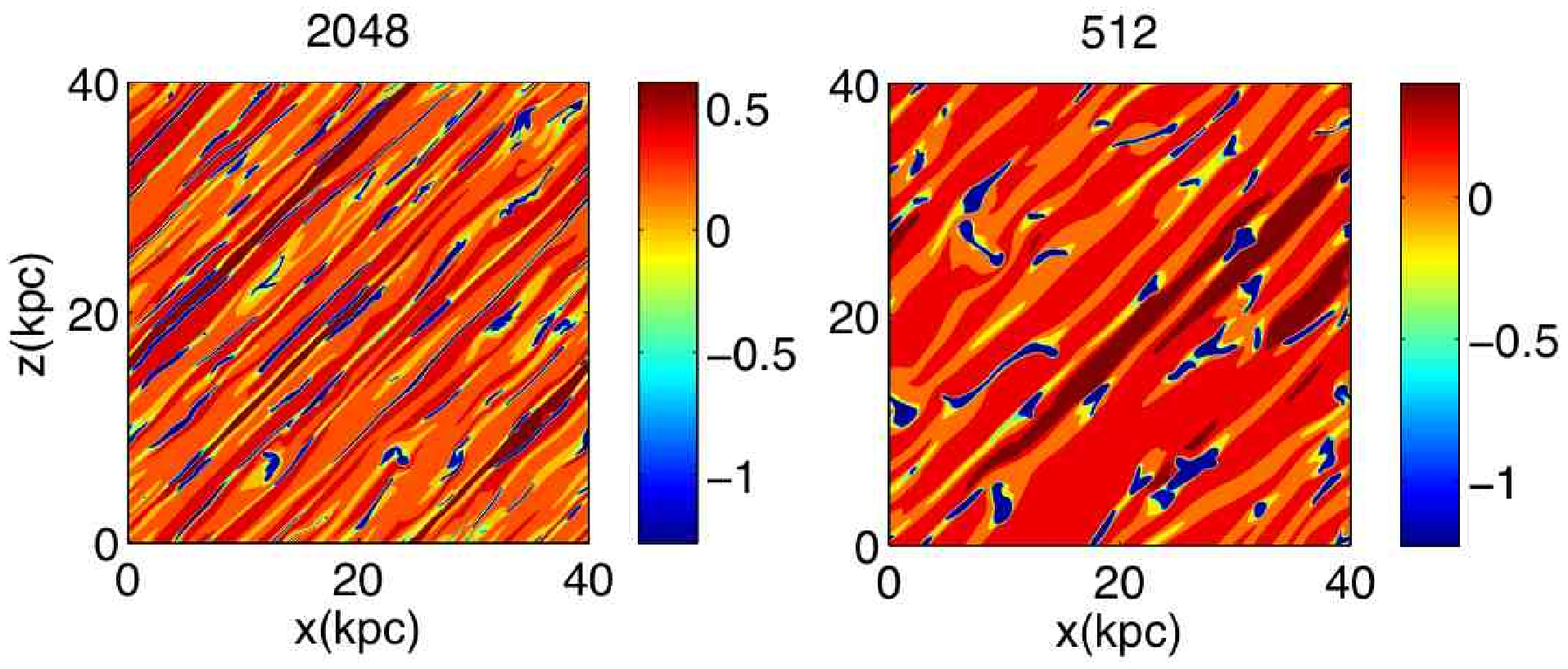}
\caption{Contour plots of $Log_{10}$ temperature (keV) at 0.95 Gyr for high (2048; {\em left}) and low (512; {\em right}) resolution simulations without the small isotropic conductivity which is needed for convergence. Compare with Figure \ref{fig:conv_2d} which shows results for simulations including a small isotropic conductivity.
  \label{fig:conv_2d_noiso}}
\end{figure}

\begin{figure}
\centering
\epsscale{1.}
\plottwo{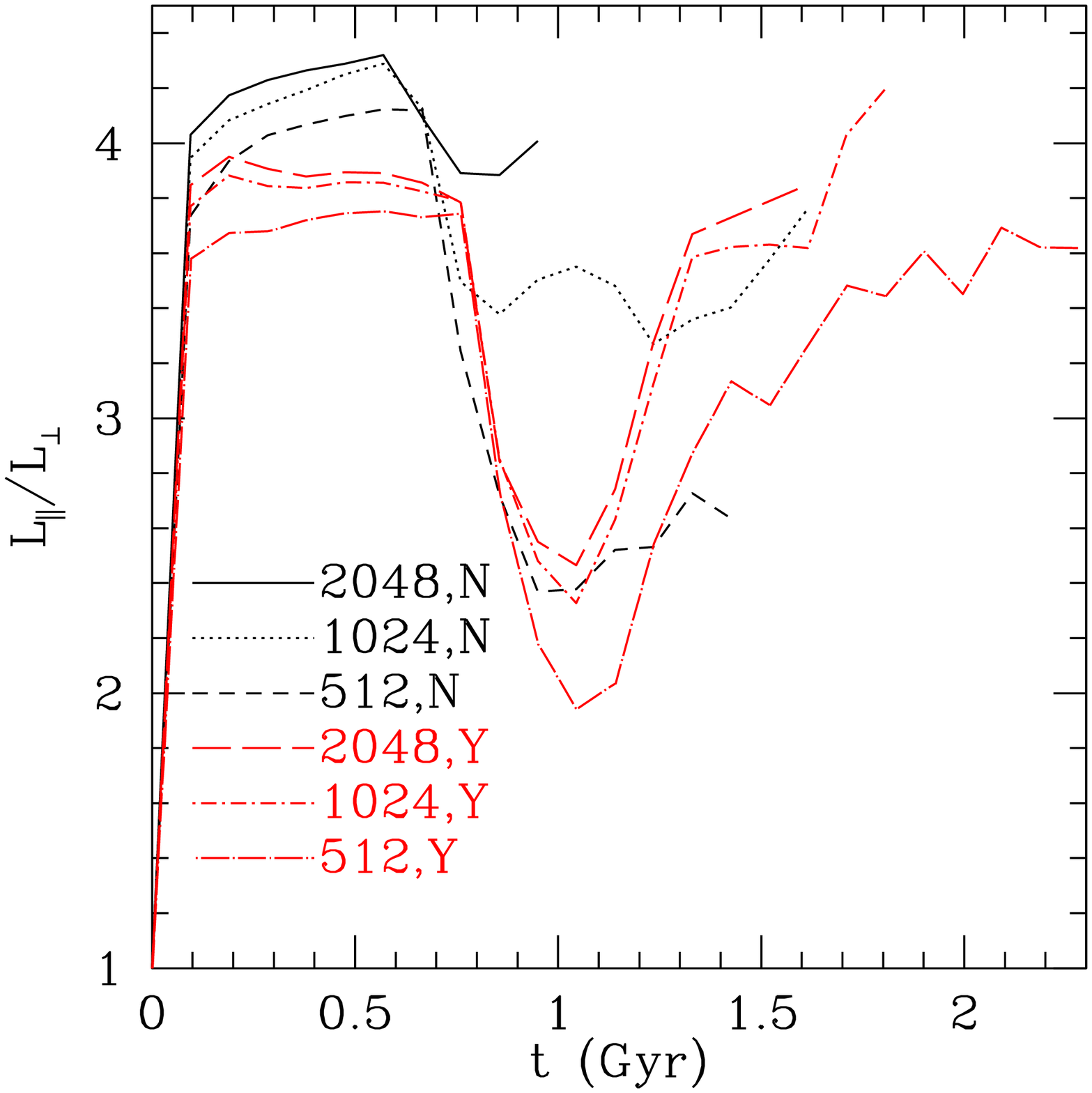}{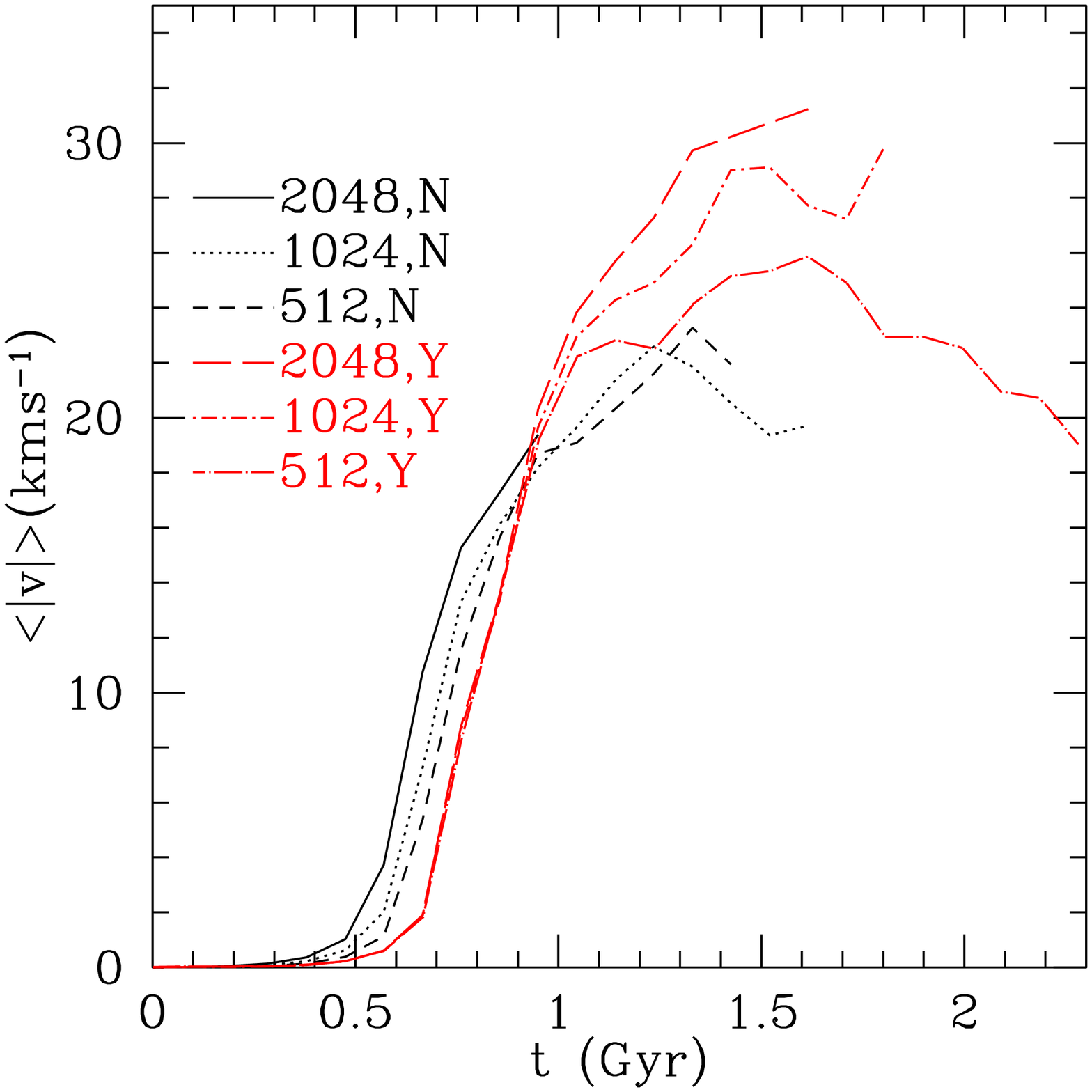}
\caption{{\em Left:} Volume averaged filament anisotropy
  ($L_\parallel/L_\perp$; Equations (\ref{eq:Lpar}) \& (\ref{eq:Lperp})) as a
  function of time.  {\em Right:} Volume averaged random velocity as a
  function of time.  In both cases, we show simulations with (labeled
  ``Y'') and without (labeled ``N'') a small isotropic conductivity
  (see \S \ref{sec:cond}).  Simulations with the isotropic conduction
  converge reasonably well with increasing resolution (for $N>512$; also see Figure \ref{fig:conv_1d}) 
  but those without it do not. We could not run the 
  higher (2048) resolution simulations for long because of very limiting 
  time-step constraints. 
  \label{fig:conv}}
\end{figure}

As described previously, in multi-dimensional simulations, the Field
length must be resolved both along and perpendicular to the
direction of the magnetic field in order for the numerical results to
converge.  Figure \ref{fig:conv_2d} shows temperature contour plots at
0.95 Gyr for runs including perpendicular conduction, with 2048 and
512 grid points, respectively. The temperature contour plots are
reasonably similar, and are similar to the results for $N = 1024$ in
Figure \ref{fig:2D}.  Figure \ref{fig:conv} provides a more
quantitative test of the convergence of the simulations: it shows that
the volume averaged values of $L_\parallel/L_\perp$ (the anisotropy of
the filaments) and $\lan |v| \ran$ (the random velocity) are almost
the same, irrespective of resolution, for runs with perpendicular
conduction (labeled ``Y'').

To explicitly illustrate the importance of including thermal
conduction perpendicular to field lines for convergence, we carried
out simulations similar to the fiducial run, but without the small
isotropic conductivity.  Figure \ref{fig:conv_2d_noiso} shows
temperature contour plots for simulations without perpendicular
conduction, for $N = $ 2048 and 512 grid points.  In this case, the
two different simulations give very different results; in particular
the filaments are much thinner and the number of filaments is much
larger (by a factor $\sim$ 4) for the higher resolution run. This is
also seen in the {left} panel of Figure \ref{fig:conv} which shows
that the anisotropy of the filaments $L_\parallel/L_\perp$ increases
with increasing resolution for runs without perpendicular
conduction. Similarly, the volume averaged velocity ($\lan |v| \ran$)
does not show clear convergence with increasing resolution in the
absence of the isotropic conductivity (see the {right} panel of
Figure \ref{fig:conv}).

It is important to stress that the simulations with perpendicular
conduction (e.g., Figure \ref{fig:conv_2d}) significantly over-estimate
the thickness of the filaments perpendicular to the magnetic field,
because the perpendicular conductivity is too large by orders of
magnitude.  In this sense the trend in Figures \ref{fig:conv} and
\ref{fig:conv_2d_noiso} is correct, namely the perpendicular
structures should indeed be thinner than in our fiducial simulations.
However, it is critical that the Field length be resolved
perpendicular to the magnetic field, or else spurious numerical
results can arise (e.g., in our simulations with cosmic rays, we found
that the cosmic ray pressure could become spuriously large in cold
filaments when they were not properly resolved, even if the initial
cosmic ray pressure was negligible). Physically, small scale turbulent
heat transport (e.g., due to Kelvin-Helmholz instabilities at the
boundaries of the filaments) or other physics (e.g., cosmic-ray
pressure; \S \ref{sec:CR}) probably sets the perpendicular scale of
the filaments, not the true microscopic perpendicular heat
transport. These processes are not currently well-understood and it is
unclear to what extent they can simply be treated as an enhanced
perpendicular conductivity (as we have done here).

\subsection{Effects of Cosmic Rays on Filament Formation}
\label{sec:CR}
\begin{figure}
\centering
\epsscale{1.}
\plotone{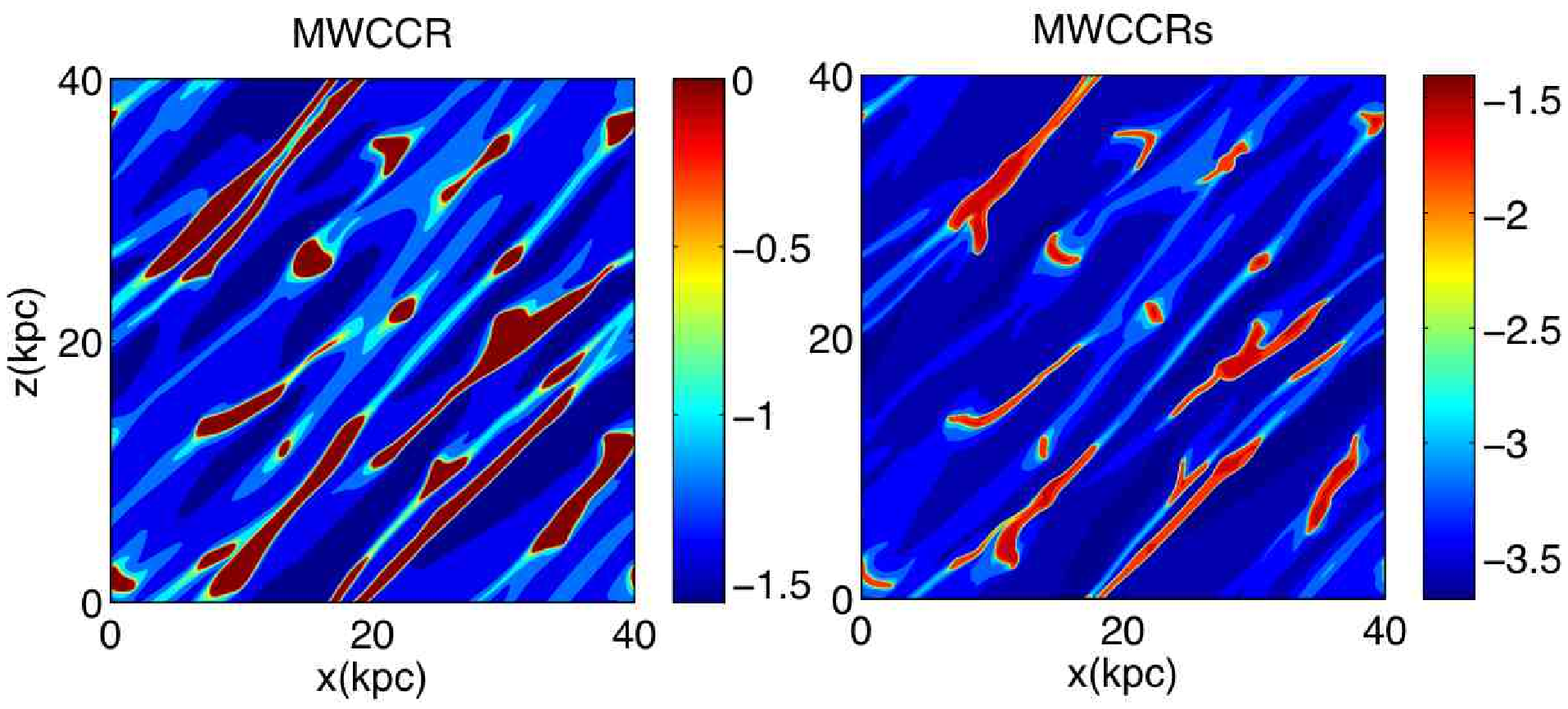}
\caption{Contour plots of the cosmic ray to plasma pressure ratio,
  $Log_{10}( p_{\rm cr}/p)$, at 0.95 Gyr for the runs with initial
  $p_{\rm cr}/p=0.1$ (MWCCR; {\em left}) and initial $p_{\rm
    cr}/p=10^{-3}$ (MWCCRs; {\em right}). The density/temperature
    contour plots look similar to these because $p_{\rm cr}/\rho^{4/3}$
    is conserved.
  \label{fig:cosmic}}
\end{figure}

In the previous sections, we have highlighted the dynamics and
thermodynamics of the thermal plasma during thermal instability.  In
this section we consider the role of cosmic-rays, i.e., a non-thermal
population of particles.  Figure \ref{fig:cosmic} shows contour plots
of the ratio of the cosmic ray to plasma pressure for simulations with
two different initial cosmic ray pressures, $p_{\rm cr}/p$ = 0.1 and
$10^{-3}$, respectively; the cosmic-rays are adiabatic in these
simulations.  Figure \ref{fig:cosmic} shows that the cosmic rays
become concentrated in the cold filaments; this is because the cosmic
ray entropy $p_{\rm cr}/\rho^{4/3}$ is conserved and the cosmic rays
are thus compressed along with the thermal plasma into the cold
filaments. For the simulations with a very small cosmic-ray pressure
({\em right} panel of Figure \ref{fig:cosmic}), the properties of the
thermal plasma in the filaments and in the diffuse ICM are very
similar to those in the simulations without cosmic rays (Figure
\ref{fig:2D}).  In particular, because the cosmic ray pressure is
small even in the nonlinear state, the cosmic rays do not affect the
physics of how the filaments form. On the other hand, when the initial
cosmic ray pressure is larger ({\em left} panel of Figure
\ref{fig:cosmic}), adiabatic compression of the cosmic rays in the
filaments leads to {\em cosmic ray pressure dominated filaments} that
are longer and broader than in the absence of cosmic rays; the
additional cosmic ray pressure halts the contraction of the filaments
when $p_{\rm cr} \sim p$.

Table \ref{tab:tab2} shows that the volume fraction $f_V$ of the cold
phase is larger for simulations in which the filaments are cosmic ray
dominated (MWCCR and MWCCRd28); the mass fraction $f_m$, however, is
smaller. This is because of the smaller gas density and thermal
pressure in the cold filaments.  In addition, Figure
\ref{fig:lbyl_comp} shows (short dashed line) 
that the filaments are more anisotropic ($L_\parallel/L_\perp$ is
larger) in the nonlinear phase for cosmic-ray dominated filaments;
this is because the cosmic ray pressure resists parallel compression.
Indeed, a visual comparison of the filaments with and without a large
cosmic-ray pressure in Figure \ref{fig:cosmic} shows that the absolute
parallel length-scale of the filaments is larger when the cosmic rays
are dynamically important.

For a realistic cooling function, the density contrast between the
filaments and the thermal plasma is much larger than in our
simulations (because the stable thermal phase has $T \simeq 10^4$ K
rather than $T \simeq 2 \times 10^6$ K).  For an initial ICM
temperature of $\sim 10^7$ K, the real density contrast should be
$\sim 10^3$ at a fixed pressure (assuming the filaments are not cosmic
ray pressure dominated).  Thus, even with an initially very small
cosmic ray pressure in the ICM of $p_{\rm cr}/p \sim 10^{-4}$, the
cosmic rays can be adiabatically compressed to be dynamically
important in filaments.  This suggests that the cosmic ray dominated
results in the {left} panel of Figure \ref{fig:cosmic} are likely to
be the most physically realistic.  However, for the large gas
densities and cosmic ray pressures that obtain in the filaments,
cosmic ray losses due to ionization, pion production, and cosmic ray
streaming will become important. The hadronic and ionization loss
timescales are comparable, $\approx 200/n_e ({\rm cm}^{-3})$ Myr, for
relativistic protons with kinetic energy of a few GeV
\citep[e.g.,][]{sch02}. The energy loss timescale because of cosmic
ray streaming is roughly the Alfv\'en crossing time along the filament
($\sim$ 1 Gyr for a 10 kpc long filament and an Alfv\'en speed of 10
km s$^{-1}$).  Since these loss timescales are only modestly longer
than the nominal cooling time, and since the filaments are expected to
be dense, cosmic ray losses have to be included
self-consistently. While including ionization and hadronic losses is
straightforward, numerically implementing cosmic ray streaming is
non-trivial (see \citealt{sha09a}).  A self-consistent treatment of
this physics is beyond the scope of the present paper, but may modify
the impact of cosmic rays on filament formation.

The only non-adiabatic cosmic ray physics in our
calculations is diffusion along magnetic field lines (Equation
(\ref{eq:crenergy})).  Our calculations with different parallel
diffusivities $D_\parallel$ show that, so long as $D_\parallel
\lesssim 10^{29}$ cm$^2$s$^{-1}$, the adiabatic results in Figure
\ref{fig:cosmic} are reasonably applicable.  \citet{sha09} presented
general arguments that the diffusivity is likely to satisfy this
inequality, so we suspect that large cosmic-ray pressures in filaments
are the norm.  There are indeed observational indications that this is
the case (e.g., the modeling of atomic and molecular lines by
\citealt{fer09}); we will discuss this comparison in \S \ref{disc}.

\subsection{Simulations with Different Cooling/Heating Functions}
\label{sec:heat}
\begin{figure}
\centering
\epsscale{1.}
\plotone{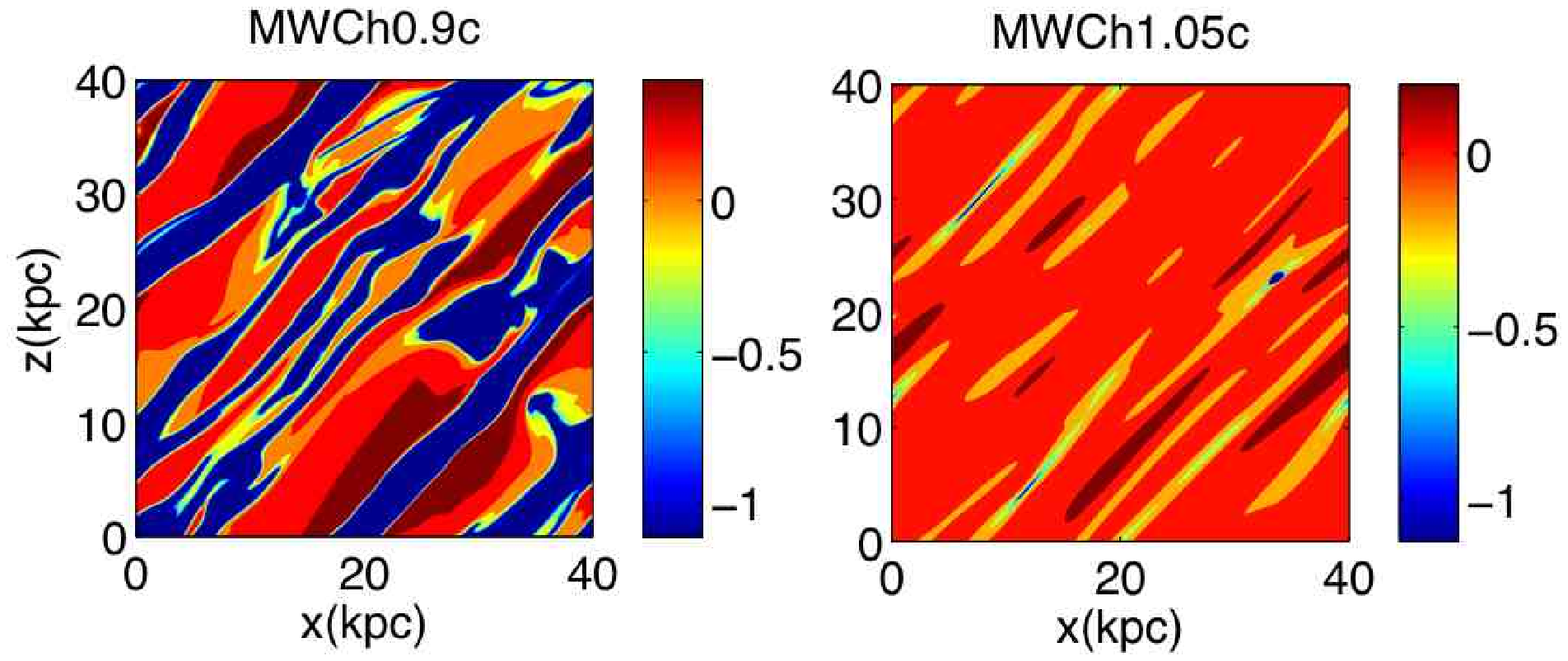}
\caption{Contour plots of $Log_{\rm 10}$ temperature (keV) at 0.95 Gyr for the run with volume averaged heating = 0.9$\times$volume averaged cooling (MWCh0.9c; {\em left}) and for the run with volume averaged heating = 1.05$\times$volume averaged cooling (MWCh1.05c; {\em right}). \label{fig:heat_not_cool}}
\end{figure}

To understand which aspects of the nonlinear evolution of the thermal
instability in the ICM are robust, we have carried out similar
calculations to those reported here with different assumptions about
the heating and/or cooling functions.  Recall that the heating
function is particularly poorly constrained in the ICM.  To give one
example, we carried out a series of simulations with the heating
proportional to density, i.e., with a heating that is constant per
unit mass instead of constant per unit volume as in our fiducial
models shown here.  The results were qualitatively similar to the
fiducial case, with anisotropic filaments and most of the volume in
the hot phase. Nonlinearly, the mass fraction in the cold phase (0.38
at 1.9 Gyr) is smaller than in the fiducial run. The cold phase is slightly hotter ($\approx 2\times 10^6$ K instead of $\approx 10^6$ K in 
the case of the fiducial run; see Figure \ref{fig:phase_time}) and has a smaller spread in temperature 
than in the fiducial run, and it takes longer for
nonlinear saturation because the cold phase is heated more effectively
than in the fiducial case. The aspect ratio of the cold filaments
(measured by $L_\parallel/L_\perp$) is similar to the fiducial run.

We also carried out simulations in which the volume averaged
instantaneous heating rate was not equal to the cooling rate: random
perturbations (up to 200\%) in both space and time were added to the
volume averaged heating rate. These runs also showed results
qualitatively similar to the fiducial run, except that the field lines
were more disturbed from the initial configuration, and the mass and
volume fractions of plasma at intermediate temperatures was larger (as
would be expected). This demonstrates that the existence of a
multiphase medium and cold filaments aligned along the magnetic field
are robust consequences of thermal instability in the ICM.  The only
way out of these conclusions is if there is a heating mechanism that
is {\em locally} thermally stable on scales $\gtrsim$ the Field
length; this is a much more stringent requirement to satisfy than the
global thermal stability of the ICM (however, see \citealt{kun10}).

The fiducial run (and all other runs) uses the modified cooling curve
shown in Figure \ref{fig:cf} with the stable phase at $T<2\times 10^6$
K. To assess what happens to filaments with a realistic cooling
function, in which the stable phase is at $<10^4$ K, we carried out two runs
(with and without cosmic rays) in which the stable phase of the
cooling curve exists for $T<10^6$ K.  The Field length in the stable
phase of these simulations is $\approx 8$ times smaller than in our
fiducial calculations (see Equation (\ref{eq:lf})). For this reason
these runs are only barely resolved (see \S\ref{sec:1d},~\ref{sec:conv} for
discussion of convergence), but they nonetheless indicate the trends
expected for a more realistic cooling function. Nonlinearly, the run with cosmic rays 
shows  much longer (and broader) filaments than the run without cosmic rays. 
This can also be seen by comparing the filaments in Figure \ref{fig:2D} (for the fiducial 
run) and Figure \ref{fig:cosmic} (for the run with cosmic rays MWCCR; 
see also Figure \ref{fig:lbyl_comp}); however, the difference is even more 
dramatic for the runs with a cooler stable phase. 
For a smaller stable phase temperature, simulations without cosmic rays have very narrow and short
filaments, while in simulations with cosmic rays the sizes of filaments do not depend significantly 
on the stable phase temperature. Thus, adiabatically compressed cosmic rays, 
which dominate the pressure in the filaments, are likely able to prevent compression of the cold 
plasma to very small scales for a realistic cooling function .

\subsection{Runs with Heating $\neq$ Cooling}
\label{sec:imbalance}

In cluster cores the instantaneous heating rate is probably not
identically equal to the cooling rate, as we have assumed in our
models. However, the inferred global stability of clusters suggests
that for timescales longer than a few cooling times heating does
roughly balance cooling. Otherwise, all of the plasma will be in the
cold phase (if cooling dominates) or in the hot phase (if heating
dominates). To test the sensitivity of the phase structure to the
degree of imbalance between heating and cooling, we carried out
simulations in which heating does not quite balance cooling (Table
\ref{tab:tab1}). Figure \ref{fig:heat_not_cool} shows temperature
contour plots after $\simeq 0.95$ Gyr for simulations with a constant
heating per unit volume = 0.9$\times$cooling (MWCh0.9c) and with
heating per unit volume =1.05$\times$cooling (MWCh1.05c). The two
plots differ dramatically. When cooling is somewhat stronger than
heating (MWCh0.9c), the filaments are longer, much broader, and
contain more of the mass, as compared to the fiducial run; by
contrast, when heating exceeds cooling (MWCh1.05c), the cold
structures are much smaller. The results differ even more dramatically
from our fiducial calculations for a larger imbalance between heating
and cooling.  When heating does not exactly balance cooling, there
will only be a single phase if we wait long enough. The relevant
timescale is $t_{\rm sec} \approx t_{\rm cool} C/|C-H|$, the timescale
for secular heating/cooling of the plasma, where $H/C$ is the volume
averaged heating/cooling rate. After $\sim t_{\rm sec}$ the plasma
will be dominated by hot/cold phase if heating/cooling dominates. Note
that Figure \ref{fig:heat_not_cool} is shown at $\simeq 0.95$ Gyr, which is
$\sim t_{\rm sec}$ for these models.  The fact that many cluster cores
show a multiphase structure implies that heating balances cooling over
a few cooling times. In the future, a more quantitative comparison
between simulations like those reported here and observations might
provide interesting constraints on the degree of thermal balance in
cluster cores.

\section{Astrophysical Implications}
\label{disc}

Early thermal stability analyses of galaxy clusters were done within
the context of the cooling flow model, in which mass inflows on the
same timescale that the plasma cools; this can significantly modify
the physics of the thermal instability
\citep[e.g.,][]{bal89}. However, observations now clearly demonstrate
the {lack} of significant cooling flows; a poorly understood source of
heating (plausibly a central AGN) roughly balances cooling, maintaining
approximate global thermal stability.  In spite of their global
stability, clusters can still be vulnerable to local thermal
instability whenever Field's criterion is satisfied (i.e., whenever
there are growing solutions to Equation (\ref{eq:isobaric})).  Thermal
conduction helps stabilize cluster plasmas on scales smaller than the
Field length (Equation (\ref{eq:lambdaF})), so it is the larger scale
perturbations that are particularly prone to instability.  It is not
guaranteed that such instabilities in fact exist: whether they do
depends on the details of how the plasma is heated. In this
paper we have used two-dimensional MHD simulations with anisotropic
conduction and cosmic rays to study the nonlinear dynamics of thermal
instability for conditions appropriate to galaxy clusters, under the
assumption that local heating is not able to maintain thermal
stability.  Our results can only be semi-quantitatively applied to
observed clusters, given current uncertainties in the heating physics.
Nonetheless, we find that none of our conclusions are that sensitive
to the precise form of the heating function (e.g., whether it is
constant per unit mass or constant per unit volume; \S
\ref{sec:heat}).  We also find similar results in simulations that
include a slight imbalance between heating and cooling (so long as the
simulation is not run too long; \S \ref{sec:imbalance}) or random
perturbations in the heating/cooling rates on top of a thermal balance
(\S \ref{sec:heat}). Observations of atomic and molecular filaments
and star formation in cool cluster cores (e.g., \citealt{cav08,ode08})
provide observational evidence for local thermally unstable regions in
clusters.

Our calculations show that, for numerical convergence, the Field
length in the {\em cold} medium needs to be resolved not only along
the magnetic field, but also perpendicular to the field lines.  To do
so, we have artificially increased the temperature at which the plasma
is thermally stable on the low temperature part of the cooling curve
(to $2 \times 10^6$ K; see Figure \ref{fig:cf}).  We have also included
a small isotropic thermal diffusivity, to ensure that the
perpendicular Field length is resolved (\S \ref{sec:cond}).

During the evolution of the thermal instability, rapid thermal
conduction along magnetic field lines suppresses compression of plasma
along the field at scales smaller than the Field length. However,
compression occurs perpendicular to field lines on large scales, where
magnetic tension is not important. Thus if the Field length is
$\lesssim$ the size of a galaxy cluster core, and if the cooling time
is short compared to the age of the cluster, the ICM is likely to be
multiphase, with atomic filaments aligned with the local magnetic
field.  Note that this conclusion holds even in the fully nonlinear
regime and does {\em not} require dynamically strong magnetic fields
(Figure \ref{fig:2D}).  Rather, thermal instability leads to a
filamentary structure because of the poor heat transport across
magnetic fields.  This result implies that the orientation of atomic
filaments can provide a local measure of the magnetic field direction
in clusters.  It also provides a physical explanation for the
filamentary structures seen in optical emission line observations of
cluster cores (\citealt{con01,spa04}).  Note that simulations with
isotropic conduction show no preference for the cold gas to align with
the magnetic field direction (e.g., Figure \ref{fig:isocond}).

The filamentary structure in the cold gas is also imprinted on the
diffuse X-ray emitting plasma in the hot ICM (e.g.,
Figure \ref{fig:2D}). Because of the large conductivity of the hot
plasma (Equation (\ref{eq:spitzer})), it is natural for a given magnetic
field line to become relatively isothermal.  If different magnetic
field lines undergo slightly different heating/cooling, as must
surely be the case to some extent, this will lead to different
temperatures, densities, and X-ray emissivities along different
magnetic field lines.
This could potentially explain the long, soft X-ray emitting
isothermal structures observed in some clusters \citep[][]{sun09}.

The ambient cluster magnetic field is enhanced by flux freezing during
the formation of filaments.  Moreover, this enhancement of B extends
over a region that is much longer than the extent of the cold gas
itself (compare Figure \ref{fig:2D_detail} for $|B|$ with Figure
\ref{fig:2D} for $T$).  This is because as the plasma compresses along
the magnetic field, it leaves behind regions devoid of much cool
plasma that have nonetheless had field amplification by flux freezing.
The volume averaged velocities induced by the thermal instability are
$\sim 25$ km s$^{-1}$ for our typical cluster parameters.  The
velocity in the hot phase is, on an average, directed toward the cold
filaments; velocities in the cold filaments are larger ($\sim 100$ km
s$^{-1}$), are generally parallel to the filaments, and may have
strong shear (because of the merger of oppositely moving filaments;
see the right panel of Figure \ref{fig:2D_detail}).  The velocities
$\sim 100$ km s$^{-1}$ we find in cold filaments are similar to the
measured random velocities of optical emission line filaments (e.g.,
\citealt{hat06}).  However, this comparison may be misleading because
we ignore gravity in our simulations which can easily induce large
motions in the dense filaments.

The length of cold filaments along the magnetic field roughly scales with the 
Field length in the cold phase. The width of filaments is also determined by the 
perpendicular Field length. For a realistic atomic cooling curve in which the cold atomic gas is at
$\sim 10^4$ K, the filaments are expected to be extremely small $\sim
10^{-4}$ pc. However, the observed atomic (e.g., H$\alpha$) filaments
are much longer than this.  This can be explained if the filaments are
supported by cosmic ray pressure which prevents the collapse of the
cold gas (see Figures \ref{fig:lbyl_comp} \& \ref{fig:cosmic}). The
presence of a significant population of cosmic rays is also inferred
by modeling the atomic and molecular lines from clusters
\citep[][]{fer09}. Even if the cosmic ray pressure is small in the
diffuse ICM ($p_{\rm cr}/p\gtrsim 10^{-4}$), adiabatic compression
will result in cosmic ray dominated cold filaments.  For the scales of
interest, cosmic ray diffusion can be neglected if the cosmic ray
diffusion coefficient is equal to the Galactic value ($10^{28}$
cm$^2$s$^{-1}$; \citealt{ber90}); \citet{sha09} argue that this is
likely to be the case.  Other loss processes (e.g., pion production,
ionization, Alfv\'en-wave excitation) may, however, be important, and
could modify how effectively cosmic rays can support filaments (\S
\ref{sec:CR}); this will be studied in more detail in future work.  If
cosmic ray pressure is indeed substantial, the pressure of the thermal
plasma in cold filaments can be significantly smaller than that of the
ambient ICM.  Substantial cosmic ray (and magnetic) pressure could in
principle help explain the {\em lack} of star formation in the
molecular filaments of NGC 1275 \citep[e.g.,][]{fab08,sal06}. Hadronic
interactions between cosmic rays and thermal nucleons in dense
filaments can produce a significant gamma-ray flux due to neutral pion
decay; however, because of the small volume occupied by the filaments,
it is unlikely that the filaments will be detectable by current
instruments.

The perpendicular thickness of the filaments in our calculations is
set largely by the isotropic diffusivity we include to ensure
numerical convergence; in reality, however, the perpendicular thermal
conductivity is negligible and some other physics (perhaps cosmic ray
pressure again) must determine the perpendicular scale of the
filaments.  The properties of observed filaments can in principle be
tested by Faraday rotation measured along and across the filaments;
the Faraday rotation should be substantial ($\sim 10^7$ rad m$^{-2}$
for $n_e \sim 10$ cm$^{-2}$, $B\sim 10\mu G$, and a filament length of
10 kpc). However, these observations are difficult precisely because
the filaments are narrow and because this requires a reasonably strong
radio source behind the filament.

We have not included gravity in our simulations in order to focus on
the thermal phase structure of the ICM and not the ICM dynamics.  In
stratified plasmas there will be a complex interplay between the
thermal instability and buoyant motions (e.g., driven by buoyancy instabilities; see 
\citealt{par08,par09a}); this will be the focus of
future work.  In the nonlinear limit, cold, dense filaments are
expected to fall, almost at the free fall rate, toward the cluster
center.  Magnetic anchoring and levitation by underdense, buoyant
bubbles may, however, prevent this (e.g., \citealt{hat06,rev08}).
Even with a significant gravitational field, we expect the filaments
to be aligned with the local magnetic field as a consequence of the
basic thermal physics of the ICM (i.e., cooling and anisotropic
thermal conduction).

\section{Acknowledgements}
Its a pleasure to acknowledge useful discussions with Chris McKee, Jim Stone, 
and particularly Eve Ostriker. Support for P.~S. and I.~J.~P. was provided by NASA through Chandra 
Postdoctoral Fellowship grant numbers PF8-90054 and 
PF7-80049 awarded by the Chandra X-ray Center, which is
operated by the Smithsonian Astrophysical Observatory for 
NASA under contract NAS8-03060. E.~Q. was supported in part by the 
David and Lucile Packard Foundation and NASA grant NNX10AC95G.
We thank the Laboratory for Computational Astrophysics, University of California, San Diego, for developing 
ZEUS-MP and providing it to the community. This research was 
supported in part by the National Science Foundation through 
TeraGrid resources provided by NCSA and Purdue University. 
The simulations reported in the paper were carried out on the 
Abe cluster at NCSA and the Steele cluster at Purdue University.

\appendix
\label{app}
\section{Linear Stability Analysis}
We assume a background hydrostatic and thermal equilibrium. Let the equilibrium quantities ($\rho_0$, $\bm{B_0}$,
$p_0$, $p_{\rm cr,0}$) be constant in space; the following analysis is valid for $kH \gg 1$ where $k$ is the wavenumber and $H$ is the scale over 
which equilibrium quantities vary. We do not include gravity in the following analysis, and hence because of $kH \gg 1$ all terms involving background
gradients are small.

Perturbations of the form $e^{(-i w t + i \bm{k}\cdot \bm{x})}$ are assumed,
where $w$ is the frequency. Linear perturbations are preceded by a $\delta$ and equilibrium quantities 
have a subscript $0$. The linearized equations are given by
\be
\frac{\delta \rho}{\rho_0} + i \bm{k}\cdot\bm{\xi}=0,
\ee
where $\bm{\xi}\equiv i \bm{\delta v}/\omega$ is the displacement vector,
\be
\label{eq:mom}
-w^2 \bm{\xi} = -i\bm{k} \frac{\delta (p + p_{\rm cr} + B^2/8\pi)}{\rho_0}  + \frac{i \bm{k} \cdot \bm{B_0}}{4 \pi \rho_0} \bm{\delta B},
\ee
\be
\label{eq:induction}
\bm{\delta B} = i (\bm{k} \cdot \bm{B_0}) \bm{\xi} - i({\bm k} \cdot \bm{\xi}) \bm{B_0},
\ee
\be
\label{eq:p}
-i w\frac{(\delta p - \gamma v_t^2 \delta \rho)}{\gamma-1}  = - \delta[n_e n_i \Lambda(T)]  - i\bm{k}\cdot \bm{\delta Q},
\ee
where $v_t^2\equiv p/\rho$, and the perturbation of space-constant $H(t)$ vanishes as it equals volume averaged cooling rate which is constant 
in time in the linear regime,
\be
\label{eq:pcr}
-i w (\delta p_{\rm cr} - \gamma_{\rm cr} v_{t,cr}^2 \delta \rho) = - i\bm{k}\cdot \bm{\delta \Gamma},
\ee
where $v_{t,cr}^2\equiv p_{\rm cr}/\rho$.

\subsection{Fast Sonic Speed Limit}
Dotting Equation (\ref{eq:mom}) with $\bm{k}$ gives $\delta p_{\rm t}/ p_{\rm t} \sim (t_{\rm snd}/t_{\rm cool})^2 \delta \rho/\rho_0$, 
where $p_{\rm t} = p + p_{\rm cr} + B^2/8\pi$, $t_{\rm snd}^{-1} \sim k (p_{\rm t}/\rho)^{1/2}$, and $w \sim t_{\rm cool}^{-1}$ (i.e., we are 
considering the condensation mode which grows at the cooling time; $t_{\rm cool}^{-1} \equiv (\gamma-1)n_en_i\Lambda/p_0$). In the limit when 
the sound-crossing time is shorter than the cooling 
time, the relative perturbation in total pressure is much smaller than the relative perturbation in density. Thus we can combine Equations
(\ref{eq:induction}), (\ref{eq:p}), and (\ref{eq:pcr}) to give,
\be
-i w \delta p_t = -i w \left (\gamma v_t^2 + \gamma_{\rm cr} v_{t,cr}^2 \right ) \delta \rho
-i w \delta \left ( \frac{B^2}{8\pi} \right ) 
- (\gamma-1) \delta[n_e n_i \Lambda(T)] - (\gamma-1) i \bm{k} \cdot \bm{\delta Q} - i\bm{k} \cdot \bm{\delta \Gamma}.
\label{eq:full}
\ee
Now in the limit of $t_{\rm snd} \ll t_{\rm cool}$, the left hand side of Equation (\ref{eq:full}) can be ignored with respect to the first term on the right 
hand side. In this limit, from Equation (\ref{eq:induction}), we get 
\be
\label{eq:delpmag}
\delta (B^2/8\pi) = v_A^2\delta \rho/(1-k_\parallel^2 v_A^2/w^2),
\ee 
where 
$v_A^2 \equiv B_0^2/4\pi \rho_0$ and $k_\parallel = \bm{k} \cdot \hat{\bm b}_0$. Thus in the short sound-crossing time limit, Equation (\ref{eq:full}) reduces to
\be
i w \left ( \gamma v_t^2 + \gamma_{\rm cr} v_{t,cr}^2  +\frac{v_A^2}{1-k_\parallel^2 v_A^2/w^2} \right ) \delta \rho
= - (\gamma-1) \delta[n_e n_i \Lambda(T)] - (\gamma-1) i \bm{k} \cdot \bm{\delta Q} - i\bm{k} \cdot \bm{\delta \Gamma},
\ee
where $p_t$ is held constant in evaluating the right hand side. The cooling term can be written as
\be
\delta[n_e n_i \Lambda(T)]_{p_t} =   \frac{\partial [n_en_i\Lambda(T)]}{\partial \rho}\vert_p \delta \rho \\   
\label{eq:dcool}
-\frac{\partial [n_en_i\Lambda(T)]}{\partial p} \vert_\rho \delta (p_{\rm cr}+B^2/8\pi),
\ee
where $\partial[n_en_i\Lambda(T)]/\partial p \vert_\rho = (n_e n_i T/p) d\Lambda/dT$, and 
$\partial[n_en_i\Lambda(T)]/\partial \rho \vert_p = -(n_e n_i T^3/\rho) d[\Lambda/T^2]/dT$. Thus, Equation (\ref{eq:dcool}), on combining with Equation (\ref{eq:delpmag}), becomes
\be
 \delta \ln [ n_e n_i \Lambda(T)]_{p_t} = - \frac{d\ln \Lambda }{d\ln T} \frac{\delta p_{\rm cr}}{p_0} 
- \left [ \frac{d  \ln (\Lambda/T^2) }{d\ln T} + \frac{2}{\beta (1-k_\parallel^2v_A^2/w^2)} \frac{d \ln \Lambda}{d\ln T} \right ] \frac{\delta \rho}{\rho_0},
\label{eq:delcool}
\ee
where $\beta\equiv 8\pi p/B^2$.
From Equation (\ref{eq:defGamma}) we obtain $i\bm{k} \cdot \bm{\delta \Gamma} = D_\parallel k_\parallel^2 \delta p_{\rm cr}$, so combining
with Equation (\ref{eq:pcr}), we get
\be
\label{eq:delpcr}
\delta p_{\rm cr} = \frac{-i w \gamma_{\rm cr} v_{t,cr}^2 \delta \rho}{(-iw + D_\parallel k_\parallel^2)}.
\ee
Similarly, $i\bm{k}\cdot\bm{\delta Q}=\kappa k_\parallel^2 \delta T$, and on using Equations (\ref{eq:p}) and (\ref{eq:delcool}), gives
\be
(-iw + \chi_\parallel k_\parallel^2) \frac{\delta p}{p_0} = t_{\rm cool}^{-1} \frac{d\ln \Lambda }{d\ln T} \frac{\delta p_{\rm cr}}{p_0} + \left (-i w \gamma + \chi_\parallel k_\parallel^2 \right .
+ \left . t_{\rm cool}^{-1} \left [ \frac{d  \ln (\Lambda/T^2) }{d\ln T} + \frac{2}{\beta (1-k_\parallel^2v_A^2/w^2)} \frac{d \ln \Lambda}{d\ln T} \right ] \right ) \frac{\delta \rho}{\rho_0},
\label{eq:lastbut1}
\ee
where $\chi_\parallel\equiv (\gamma-1) \kappa_\parallel T/p$ is the thermal diffusivity. Combining Equations
(\ref{eq:lastbut1}), (\ref{eq:delpcr}), and (\ref{eq:delpmag}), and using $\delta p_t\approx 0$ gives,
\ba
\nonumber
&& \frac{-i w \alpha}{(-i w+D_\parallel k_\parallel^2)}\left( -i w +\chi_\parallel k_\parallel^2 + t_{\rm cool}^{-1} \frac{d\ln \Lambda }{d\ln T}  \right) 
- \frac{2}{\beta}
\frac{(-iw +\chi_\parallel k_\parallel^2) }{(1-k_\parallel^2v_A^2/w^2)} = - i w \gamma + \chi_\parallel k_\parallel^2  \\
\label{eq:disp}
&+&  t_{\rm cool}^{-1} \left [ \frac{d  \ln (\Lambda/T^2) }{d\ln T} + \frac{2}{\beta (1-k_\parallel^2v_A^2/w^2)} \frac{d \ln \Lambda}{d\ln T} \right ], 
\ea
where $\alpha=p_{\rm cr}/p$.
In the hydro limit ($\alpha \ll 1$ and $\beta \gg 1$, irrespective of $k_\parallel^2v_A^2 t_{\rm cool}^{-2}$; i.e., magnetic tension plays no role in the condensation mode), we recover the classic isobaric thermal instability stabilized by conduction along field lines (Equation (\ref{eq:isobaric})). 
For $\beta \ll 1$ or $\alpha \gg 1$ one obtains the thermal instability in the isochoric limit (see next section), with conductive 
stabilization for scales smaller than the isochoric Field length (Equation (\ref{eq:isochoric})).
The condensation mode is isochoric when magnetic/cosmic ray pressure dominates because the constancy of the total pressure is equivalent to 
the constancy of the magnetic/cosmic ray pressure, and from Equations (\ref{eq:delpmag}) and (\ref{eq:delpcr}) a constant magnetic/cosmic ray 
pressure implies a constant density.

\subsection{Slow Sonic Speed Limit}
In the opposite limit, $t_{\rm cool}\ll t_{\rm snd}$, $\delta \rho/\rho \ll \delta p_{\rm t}/p_{\rm t}$, Equation (\ref{eq:full}) gives
\be
-iw \delta p_t =   - (\gamma-1) \delta[n_e n_i \Lambda(T)] - (\gamma-1) {i \bm{k} \cdot \bm{\delta Q}}- i\bm{k} \cdot \bm{\delta \Gamma},
\ee
where terms on the right hand side are evaluated keeping the density constant (i.e., isochoric). The perturbed magnetic and cosmic ray pressure
vanish in the isochoric limit (Equations (\ref{eq:delpmag}), (\ref{eq:delpcr})) and $\delta p_t = \delta p$, and the dispersion relation is the same as Equation (\ref{eq:isochoric}).

\end{document}